\documentclass[%
 prt,
 reprint,
 superscriptaddress,
 preprintnumbers,
 nofootinbib,
 amsmath,amssymb,
 aps,
]{revtex4-1}

\usepackage{graphicx}
\usepackage{dcolumn}
\usepackage{color}
\usepackage[colorlinks=true,citecolor=blue,urlcolor=blue]{hyperref}
\usepackage{bm}
\usepackage{todonotes}

\begin{document}

\title{Self-Interacting Neutrinos in Light of Large-Scale Structure Data}

\author{Adam He}
\affiliation{%
 Department of Physics and Astronomy, University of Southern California, Los Angeles, CA 90089, USA}
\author{Rui An}
\affiliation{%
 Department of Physics and Astronomy, University of Southern California, Los Angeles, CA 90089, USA}
\author{Mikhail M. Ivanov}
\affiliation{Center for Theoretical Physics, Massachusetts Institute of Technology, Cambridge, MA 02139, USA}
\author{Vera Gluscevic}
\affiliation{%
 Department of Physics and Astronomy, University of Southern California, Los Angeles, CA 90089, USA}

\begin{abstract}
We explore a self-interacting neutrino cosmology in which neutrinos experience a delayed onset of free-streaming. We use the effective field theory of large-scale structure (LSS) to model matter distribution on mildly non-linear scales within the self-interacting neutrino cosmology for the first time. We perform the first combined likelihood analysis of BOSS full-shape galaxy clustering, weak lensing, and Lyman-$\alpha$ forest measurements, together with the cosmic microwave background (CMB) data from \textit{Planck}. We find that the full data set strongly favors presence of a flavor-universal 
neutrino self-interaction, with a characteristic energy 
scale of order $10$~MeV. The preference is at the $\sim 5\sigma$ level and is primarily driven by the Lyman-$\alpha$ forest measurements and, to a lesser extent, the weak lensing data from DES. 
The self-interacting neutrino model eases both the Hubble tension and the $S_8$ tension between different cosmological data sets, but it does not resolve either. Finally, we note a preference for a non-zero sum of neutrino masses at the level of $\sim 0.3$~eV under this model, consistent with previous bounds. 
These results call for further investigation in several directions, and may have significant implications for neutrino physics and for future new-physics searches with galaxy surveys.

\end{abstract}

\preprint{MIT-CTP/5608}
\maketitle

\section{\label{sec:level1}Introduction\protect}

Since the discovery of neutrino oscillations and the resulting implication that neutrinos are \textit{not} massless, the neutrino sector of the Standard Model (SM) has received renewed attention as a potential window into physics beyond the SM. 
In particular, the existence of neutrino mass may imply that neutrinos experience additional couplings to particles that have not yet been observed~\citep{P_s_2002}.
While there are a number of terrestrial experiments capable of probing neutrino interactions \citep{Aguilar_Arevalo_2021, icecubecollaboration2023measurement, novacollaboration2023measurement, Adamson_2020}, cosmological observations also offer a rich landscape of complementary information that may be used to explore and constrain the neutrino sector. 
For instance, cosmic microwave background (CMB) data from the $Planck$ satellite and baryon acoustic oscillation (BAO) measurements from galaxy redshift surveys have placed a competitive upper limit on the sum of neutrino masses $\sum m_\nu < 0.12$ eV \citep{Planck2018_VI}. 
Moreover, \cite{Choi_2018, Song_2018, Lorenz_2019, Barenboim_2019, Forastieri_2019, Smirnov_2019, Escudero_2020, Ghosh_2020, Funcke_2020, Sakstein_2020, Mazumdar_2020, Blinov_2020, de_Gouv_a_2020, Froustey_2020, Babu_2020, Deppisch_2020, Kelly_2020, Abenza_2020, He_2020, Ding_2020, Berbig_2020, Gogoi_2021, Barenboim_2021, Das_2021, Mazumdar_2022, Brinckmann_2021, Kelly_2021, Esteban_2021, Arias_Arag_n_2021, Du_2021, Carrillo_Gonz_lez_2021, Huang_2021, Sung_2021, Escudero_2021, Choudhury_2021, Roy_Choudhury_2022, Carpio_2023, Orlofsky_2021, Esteban_2021_2, Venzor_2022, Venzor_2023} have found that cosmological data are capable of detecting new interactions in the neutrino sector, particularly those that would change the free-streaming nature of neutrinos after they decouple from the SM.

Cosmological studies have focused on parameterizing the neutrino self-interaction rate as $\Gamma_{\nu} \propto G_\mathrm{eff}^2T_{\nu}^5$, where $G_\mathrm{eff}$ is a Fermi-like coupling constant that describes a four-fermion interaction, and $T_{\nu}$ is the background temperature of the neutrinos. 
An example model that would produce such an interaction rate is neutrino self-scattering through a massive scalar particle $\phi$ \citep{SINu_white_paper}. 
In the presence of self-interactions, neutrino free-streaming is delayed, which leaves an imprint on matter clustering in the early and late universe. In particular, neutrino self-scattering reduces the size of the sound horizon at recombination, increasing the value of $H_0$ \citep{Knox:2019rjx}. Related to this effect, the net phase shift in the CMB acoustic peaks produced by neutrino free-streaming in standard cosmology \citep{Bashinsky_2004,Baumann_2016,Follin_2015} is absent if neutrinos feature self-scattering~\citep{Kreisch_2020}. Finally, the self-interactions suppress power and alter the linear matter power spectrum $P(k)$ at small scales in a scale--dependent manner, testable by a variety of cosmological probes; see Fig.~\ref{fig:power spectra}.

\cite{Kreisch_2020} and \cite{Kreisch_2022} have found that a significant self-interaction between neutrinos provides a good fit to CMB data from the \textit{Planck} and ACT experiments. More precisely, the CMB anisotropy is consistent both with no interaction, as well as with a sizeable value of the self-coupling constant $G_\mathrm{eff}$ \citep{Cyr_Racine_2014, Archidiacono_2014, Lancaster_2017, Oldengott_2017}. 
In other words, when analyzing only CMB data, the 1D marginalized posterior probability distribution for the $G_\mathrm{eff}$ parameter is found to be bimodal, featuring a high-probability mode consistent with $\Lambda$CDM cosmology (previously dubbed ``moderately-interacting mode'' MI$\nu$ \citep{Kreisch_2020} because of its non-zero best-fit value of $G_\mathrm{eff}$), and a ``strongly-interacting mode'' SI$\nu$, at $G_\mathrm{eff}\sim 0.03\ \mathrm{MeV}^{-2}$.
In addition, the best-fit cosmology with a large neutrino self-coupling was shown in \cite{Kreisch_2020} to be consistent with an increased value of the Hubble parameter $H_0$ in the CMB analysis, which alleviates the Hubble tension between the CMB and the late-time measurements of the expansion rate. On the other hand, posteriors analyses of \textit{Planck} data in \cite{Brinckmann_2021} indicated that a strong neutrino self-coupling could not ease the Hubble tension any more than a variable $N_\mathrm{eff}$ parameter. However, while numerous beyond-cold-dark-matter (beyond-CDM) models have been shown to alleviate cosmological tensions, few have been able to simultaneously address the Hubble tension and the $S_8$ tension---the mild discrepancy in the late-time amplitude of matter perturbations $S_8$, as inferred from the CMB and from the large-scale structure (LSS) probes \citep{Abdalla_2022}. Self-interacting neutrinos are a potential solution to both tensions~\citep{Kreisch_2020}. 

Previous analyses of the interacting-neutrino model have treated the CMB measurements alone. In this work, uniquely enabled by employing the effective field theory (EFT) of LSS \citep{Baumann:2010tm,Carrasco:2012cv,Cabass:2022avo,Ivanov:2022mrd}, we present the first joint analysis of the bulk of available LSS data in the context of the self-interacting neutrino model, together with the CMB anisotropy from \textit{Planck}.\footnote{While working on this paper, we became aware of work~\citep{camarena2023twomode} that carried out an analysis of the BOSS full-shape power spectrum in the context of self-interacting neutrinos, without considering the CMB. When they overlap, our results agree.} Specifically, we use data from the Baryon Oscillation Spectroscopic Survey (BOSS), Lyman-$\alpha$ forest measurements from the Sloan Digital Sky Survey (SDSS eBOSS), and weak lensing measurements from the Dark Energy Survey (DES). We combine the LSS measurements with \textit{Planck} measurements of the CMB primary and lensing anisotropy, modeling all CMB and LSS observables self-consistently. 

To capture cosmological effects of massive self-interacting neutrinos, our key analyses include the self-coupling constant $G_\mathrm{eff}$ and the sum of the neutrino masses $\sum m_\nu$, in addition to the six standard $\Lambda$CDM cosmological parameters. Taking all data sets at face-value, we find that a delayed onset of neutrino free-streaming consistent with a flavor-universal neutrino self-interaction is preferred to the $\Lambda$CDM cosmological model, providing a good fit to all LSS data and to the \textit{Planck} temperature, polarization, and lensing anisotropy measurements. The best-fit coupling constant is $G_\mathrm{eff}\sim (10~\mathrm{MeV})^{-2}$, consistent with the SI$\nu$ mode reported in~\cite{Kreisch_2022}. The strong-coupling mode of the posterior is favored over zero-coupling when DES data are included in the analysis; with the addition of Lyman-$\alpha$ forest data, the posterior retains only the strongly-coupled neutrino self-interaction mode. Consistent with the previous CMB-only analysis \cite{Kreisch_2020}, we find that neutrino self-scattering marginally eases---though does not resolve---cosmological tensions. Finally, we find that the best-fit cosmology features a non-vanishing sum of neutrino masses, both within $\Lambda$CDM and after including neutrino self-interactions. 
These results could have important implications for neutrino physics and for the concordance model of cosmology, and they warrant further detailed investigation with existing and upcoming data sets. 

This paper is organized as follows. In Sec.~\ref{sec:formalism}, we describe the neutrino cosmology and the EFT of LSS in the context of neutrino self-scattering. 
We describe the choice of data in Sec.~\ref{sec:data} and the analysis in Sec.~\ref{sec:analysis}.
In Sec.~\ref{sec:results}, we present our results. We discuss their implications and conclude in Sec.~\ref{sec:discussion}.

\section{\label{sec:formalism}Distribution of matter\protect}

In a self-interacting neutrino cosmology, the Boltzmann equations for the massive neutrino multipoles $\nu_\ell$ contain additional collision terms that account for neutrino self-scattering~\citep{Kreisch_2020}. In synchronous gauge, the Boltzmann equations are

\begin{equation}\label{boltzmann}
\begin{split}
    \frac{\partial \nu_\ell}{\partial \tau} + k\frac{q}{\epsilon}\left(\frac{\ell+1}{2\ell+1}\nu_{\ell+1}-\frac{\ell}{2\ell+1}\nu_{\ell-1}\right)& \\
    +\frac{2}{3}\left(\frac{\partial h}{\partial \tau}\delta_{\ell0}-\frac{2}{5}\left(\frac{\partial h}{\partial \tau}+6\frac{\partial \eta}{\partial \tau}\right)\delta_{\ell2}\right)& \\
    = - \frac{a G_{\mathrm{eff}}^2 T_{\nu}^5 \nu_{\ell}}{f_{\nu}^{(0)}(q)}\left(\frac{T_{\nu,0}}{q}\right)
    \left(A \left( \frac{q}{T_{\nu,0}}\right)\right.& \\
    \left.+B_{\ell} \left(\frac{q}{T_{\nu,0}}\right)-2D_{\ell} \left(\frac{q}{T_{\nu,0}}\right)\right)&
\end{split}
\end{equation}

\noindent where $q$ is the magnitude of the comoving momentum, $\epsilon=\sqrt{q^2 + a^2 m_\nu^2}$, $f_{\nu}^{(0)}$ is the background neutrino distribution function, $T_{\nu,0}$ is the present-day temperature of the neutrinos, $h$ and $\eta$ are the standard metric perturbation variables, and $A(x)$, $B_\ell(x)$, and $D_\ell(x)$ are functions that capture the collision term at first order in perturbation theory. In the massless case, the neutrino multipole hierarchy simplifies to

\begin{equation}
\begin{split}
    \frac{\partial F_\ell}{\partial \tau} + k\left(\frac{\ell+1}{2\ell+1}F_{\ell+1}-\frac{\ell}{2\ell+1}F_{\ell-1}\right)&\\
    +\frac{2}{3}\left(\frac{\partial h}{\partial \tau}\delta_{\ell0}-\frac{2}{5}\left(\frac{\partial h}{\partial \tau}+6\frac{\partial \eta}{\partial \tau}\right)\delta_{\ell2}\right)& \\
    = -a G_\mathrm{eff}^2 T_{\nu}^5 \alpha_\ell F_\ell&
\end{split}
\end{equation}

\noindent where $F_\ell$ are the massless neutrino multipoles and $\alpha_\ell$ is a coefficient calculated from an integral over $A(x)$, $B_\ell(x)$, and $D_\ell(x)$. See \cite{Kreisch_2020} for a detailed description of the collision terms $A(x)$, $B_\ell(x)$, $D_\ell(x)$, and $\alpha_\ell$.

We configure the Boltzmann solver \texttt{CLASS} \citep{Brinckmann_2018} to allow for neutrino self-interactions as given by these modified Boltzmann equations. Following \cite{Kreisch_2020}, we precompute $A$, $B_\ell$, and $D_\ell$ on a 5-point grid of $q/T_{\nu,0}$ values and access them via an interpolation routine as the equations are solved; similarly, we precompute $\alpha_\ell$ for the case of massless neutrinos. Also following \cite{Kreisch_2020}, we assume that the neutrino sector is comprised of one massive neutrino and allow the remaining number of massless neutrino species to vary. We implement a tight-coupling approximation in which multipoles $\ell\geq2$ are set to zero until the neutrino self-interaction rate falls below $1000*aH$, where $H$ is the Hubble parameter.
\begin{figure}[t!]
\includegraphics[scale=0.53]{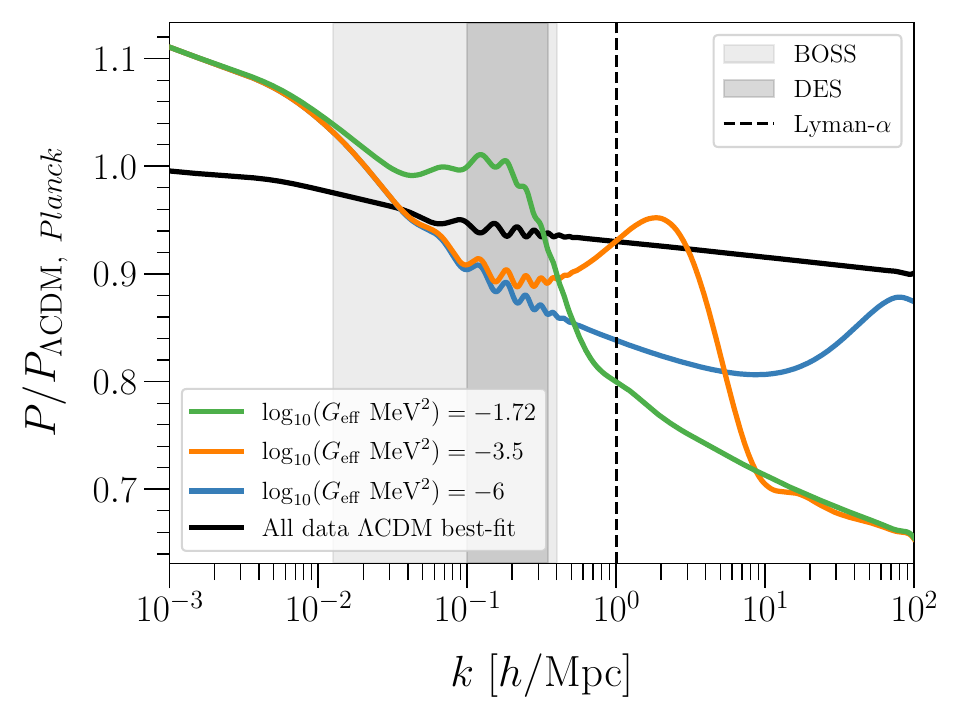}
\caption
{Ratio of the linear matter power spectra for a self-interacting neutrino cosmology and $\Lambda$CDM cosmology, using best-fit parameters from an analysis of the entire data set (\textit{Planck}+LSS).
The three colored curves differ only in the value of the self-coupling constant $G_\mathrm{eff}$: green represents the best-fit value, while the orange and the blue correspond to weaker couplings. The black curve corresponds to the best-fit parameters from an all-data analysis of the $\Lambda$CDM+$\sum m_\nu$ model. All curves are divided by the best-fit power spectrum from a \textit{Planck}--only analysis of $\Lambda$CDM. The light and the dark shaded regions roughly indicate the ranges of scales probed by BOSS and DES, respectively, and the dashed vertical line approximately corresponds to the Lyman-$\alpha$ forest measurements. An increase in $G_\mathrm{eff}$ shifts the bump-like feature towards larger physical scales and is associated with a later onset of neutrino free-streaming.
\label{fig:power spectra}}
\end{figure}

We merge this modified \texttt{CLASS} code with \texttt{CLASS-PT}~\citep{Chudaykin2020}\footnote{\url{https://github.com/Michalychforever/CLASS-PT}}, which is tailored to compute LSS power spectra in the mildly non-linear regime. \texttt{CLASS-PT} calculates non-linear 1-loop corrections to the linear matter power spectrum, and outputs the redshift-space galaxy power spectrum. \texttt{CLASS-PT} uses the EFT of LSS~\citep{Baumann:2010tm,Carrasco:2012cv,Cabass:2022avo,Ivanov:2022mrd} 
to model the redshift-space galaxy power spectrum from $0.01<k<0.2 \ h/\mathrm{Mpc}$; in the context of neutrino self-interactions, the EFT should in principle be modified to account for such a scenario.
However, non-linear effects are entirely negligible at the high redshifts where neutrino self-interactions shape the evolution of matter perturbations.
At the same time, neutrino self-interactions are effectively halted at the lower redshifts relevant for galaxy surveys, where matter distribution evolves as in $\Lambda$CDM, but with a modified initial power spectrum; see Fig.~\ref{fig:power spectra}.
Thus, the standard version of \texttt{CLASS-PT} is apt for predicting late-time LSS observables in the context of self-interacting neutrinos.\footnote{Note that typical neutrino masses in our chains go up to $m_\nu\sim 0.5~$eV, for which  
corrections beyond the current \texttt{CLASS-PT} implementation \citep{Ivanov:2019hqk} may be marginally important, see~\cite{Chiang:2018laa,Garny:2022fsh}. We leave their detailed incorporation along the lines of \cite{Senatore_2017}
for future work.} 

\begin{figure}[t!]
\includegraphics[scale=0.54]{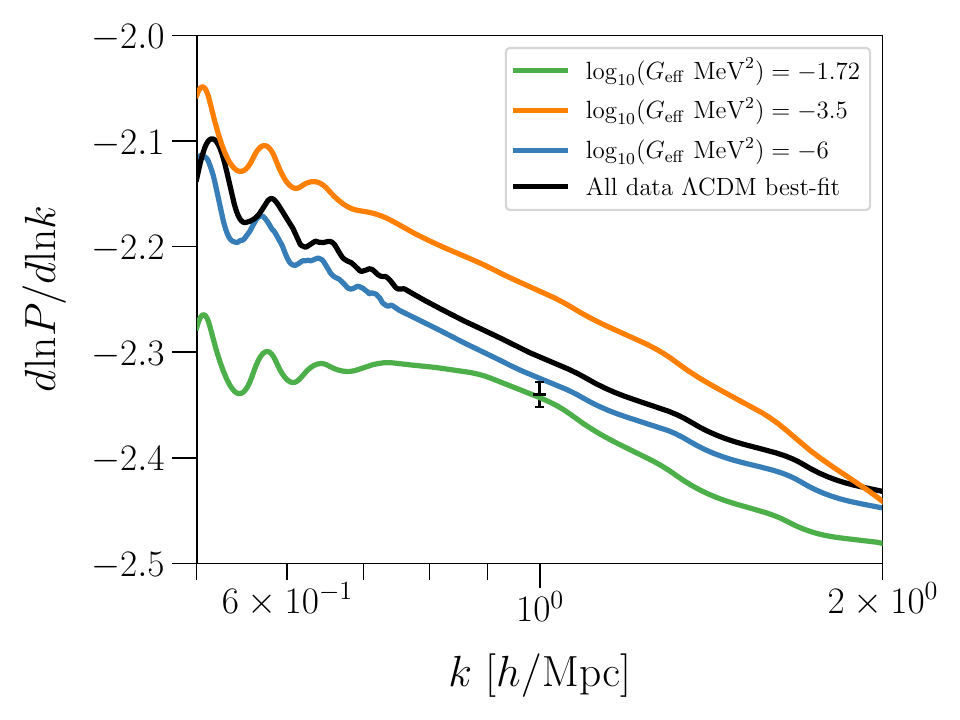}
\caption
{The slope of the linear matter power spectrum is shown for a self-interacting neutrino cosmology and $\Lambda$CDM, generated with best-fit parameter values from an analysis of the entire data set (\textit{Planck}+LSS). The parameters for the self-interacting cases are chosen in the same way as in Fig.~\ref{fig:power spectra}.
The data point indicates the slope measurement derived from the Lyman-$\alpha$ forest, with a $2 \sigma$ uncertainty.
\label{fig:derivative}}
\end{figure}

We display the effect of neutrino self-interactions on the matter power spectrum in Fig.~\ref{fig:power spectra} and on its slope in Fig.~\ref{fig:derivative}. In both cases, we use the \textit{Planck}--only best-fit cosmology for our $\Lambda$CDM calculations; for the self-interacting neutrino cosmology, we use the best-fit parameters derived from the entire data set, with LSS included. The three colored curves only differ by the value of $G_\mathrm{eff}$, illustrating the effect of the increase in the self-coupling on cosmological perturbations. We note that the green curve is preferred by the data, while the other two (with weaker interaction strengths) are disfavored. We also note that the power suppression resulting from self-interactions is progressively more pronounced on smaller scales. At the same time, the scale of the particle horizon at neutrino decoupling shows up as a bump-like feature prior to the onset of power suppression, altering both the slope and the amplitude of the power spectrum in a scale-dependent manner; this feature occurs at larger $k$ values for weaker self-interactions. All of these effects combined produce a characteristic $k$--dependent alteration to $P(k)$.

\section{\label{sec:data}Data\protect}

We analyze a combination of the following data sets:

\begin{itemize}
    \item \textbf{\textit{Planck}}: \textit{Planck} 2018 CMB \texttt{plik\_lite} high-$\ell$ TT/TE/EE likelihood, along with the \texttt{commander} low-$\ell$ TT likelihood and the \texttt{smica} lensing likelihood \citep{Planck2018_V}.
    \item \textbf{BOSS}: anisotropic galaxy clustering data from BOSS DR12 at $z=0.38$ and 0.61 \citep{Alam_2017,Ivanov_2020,Ivanov:2019hqk}. As in~\cite{Chudaykin:2020ghx,Philcox:2021kcw}, our analysis is performed up to $k_\mathrm{max} = $ 0.2~$h/\mathrm{Mpc}$ for the galaxy power spectrum multipoles, 
    from $ 0.2<k<  0.4$~$h/\mathrm{Mpc}$ for the real-space
    power spectrum proxy $Q_0$~\citep{Ivanov:2021fbu}, and up to 
    $k_\mathrm{max} = $ 0.08~$h/\mathrm{Mpc}$
    for the bispectrum monopole~\citep{Ivanov:2021kcd,Philcox:2021kcw}.\footnote{The BOSS full-shape likelihood that we use is 
    available at~\mbox{\url{https://github.com/oliverphilcox/full_shape_likelihoods}},
    see~\cite{Philcox:2021kcw} for more detail.
Also see~\cite{Chen:2021wdi,Zhang:2021yna} for alternative but equivalent likelihoods.}
    We also add post-reconstructed BOSS DR12 BAO data following~\citep{Philcox:2020vvt}.\footnote{Additional LSS data, e.g. power 
    spectra of eBOSS emission line 
    galaxies (ELG)~\citep{Ivanov:2021zmi} and quasars (QSO)~\citep{Chudaykin:2022nru}, as well as 
    BOSS bispectrum multipoles~\citep{Ivanov:2023qzb}, do 
    not sharpen cosmological constraints significantly. 
    Besides, the eBOSS QSO and ELG 
    measurements are noticeably affected by 
    observational systematics, which 
    complicate their interpretation. 
    Given these reasons, we do not 
    include these data sets in this analysis.}
    \item \textbf{Lyman-$\alpha$}: 1D Lyman-$\alpha$ flux power spectrum from SDSS DR14 BOSS and eBOSS quasars~\citep{Chabanier:2018rga}.
    We use the compressed version of this 
    likelihood presented as a Gaussian 
    prior on the 
    model-independent amplitude and slope of the power spectrum at a pivot redshift $z_p=3$ and wavenumber $k_p=0.009 \ \mathrm{s/km} \sim 1 \ h/\mathrm{Mpc}$~\citep{goldstein2023canonical}. 
    \item \textbf{DES}: weak lensing data from Year 3 of the Dark Energy Survey (DES-Y3), in the form of a Gaussian prior on $S_8$: $0.776\pm0.017$~\citep{Abbott_2022}.
    
\end{itemize}

In our EFT-based full-shape analysis, we consistently marginalize over all necessary 
nuisance parameters that capture galaxy bias, baryonic feedback, non-linear redshift space-distortions, etc.~\citep{Philcox:2021kcw}.\footnote{Our priors 
are significantly wide to 
ensure that our main cosmological 
results are unbiased (cf.~\cite{Nishimichi:2020tvu}), and not driven by the priors 
on nuisance parameters. We also note that our priors are motivated
directly by the physics of BOSS red luminous galaxies 
\citep{Chudaykin:2020ghx}, 
and are fully consistent with the 
posteriors.
See~\cite{Zhang:2021yna}
for another prior choice.  
}
Our analysis is thus robust to the specific details of galaxy formation physics. We also apply our BOSS galaxy clustering data to the self-interacting neutrino scenario without any $\Lambda$CDM assumptions; namely,
we do not use the ``compressed'' BOSS likelihood containing BAO and RSD parameters that are derived with a fixed \textit{Planck}-like $\Lambda$CDM template~\citep{Ivanov_2020,eBOSS:2020yzd}. As in \cite{Ivanov_2020}, our EFT-based 
likelihood includes galaxy power spectrum shape information that the standard BOSS
likelihood does not contain \citep{Alam_2017}. Finally, we verify that there are no degeneracies of the EFT bias parameters with any of our physical parameters of interest.

The choice we make to impose a DES prior on $S_8$ is equivalent to adding the complete DES-Y3 dataset to our analysis, as DES measures $S_8$ to be the same value for $\Lambda$CDM, WDM, and $\Lambda$CDM extensions \citep{Abbott_2022, DES_extensions_2021, Hill_2020,Ivanov:2020ril}. The value of $S_8$ is therefore independent of the choice of the cosmological model, as long as the late-time growth of structure is not modified; this is indeed the case for the interacting-neutrino model. Moreover, $S_8$ is the primary directly 
observed principle component of the weak lensing data; it is thus close to being a model-independent quantity. Therefore, we safely leave details of the full calculation of the DES-Y3 likelihood in context of neutrino self-interactions for future work.

Finally, we note that the data we chose to analyze, including BOSS, Lyman-$\alpha$, and DES, are good proxies for the information gleaned from LSS, but they do not represent the complete set of data currently available. In particular, we did not consider weak lensing from KiDS-1000 and HSC-Y3 because they have non-negligible covariance with the data sets we consider; this covariance is not yet available and must be modeled to analyze all these data in tandem \citep{DES_2022}. This task is beyond the scope of our work. 
Furthermore, we do not analyze supernovae data from Pantheon+ \citep{Brout_2022}, since this data would solely constrain $\Omega_\mathrm{m}$, a background quantity that does not change from its $\Lambda$CDM value under neutrino self-interactions. 
Finally, we do not include eBOSS DR16 BAO data in our analysis, but we expect this data to further prefer a delayed onset of neutrino free-streaming, as was the case in previous analyses \citep{Kreisch_2020,Kreisch_2022}.
Likewise, eBOSS DR16 has not yet been fully converted into a full-shape likelihood, so we do not include it here.
A dedicated future study is warranted to analyze together all these data sets under cosmological models beyond $\Lambda$CDM.

\section{\label{sec:analysis}Analysis\protect}

We use a modified \texttt{CLASS} code and \texttt{MontePython} and \texttt{MultiNest} samplers to perform likelihood analyses \citep{Brinckmann_2018, Audren_2012, Feroz_2008}. We assume that the neutrino sector consists of one massive neutrino and that the rest of the species are massless (see Sec.~\ref{sec:formalism}).\footnote{We do not expect this assumption to affect our results, as confirmed by \cite{Kreisch_2020, Kreisch_2022}.} We use the standard Big Bang nucleosynthesis (BBN) predictions for the primordial Helium abundance $Y_\mathrm{He}$, following~\cite{Kreisch_2020,Kreisch_2022}. 

In each posterior analysis, we vary all standard cosmological parameters: the baryon density $\omega_{\mathrm{b}}$, DM density $\omega_{\mathrm{c}}$, angular size of the sound horizon $\theta_\mathrm{s}$, optical depth to reionization $\tau_\mathrm{reio}$, fluctuation amplitude $A_{\mathrm{s}}$, and spectral index $n_{\mathrm{s}}$. For the interacting-neutrino cosmology, we set the effective self-coupling constant $G_\mathrm{eff}$ and the sum of the neutrino masses $\sum m_\nu$ as additional free parameters of the fit; we label this as our `baseline' model, $G_\mathrm{eff}$+$\sum m_\nu$. 
Unless otherwise noted, we fix $N_\mathrm{eff}=3.046$ to its standard--model value in our runs. 
Analogously, we consider an extended--$\Lambda$CDM model where we vary $\sum m_\nu$ and label it as $\Lambda$CDM+$\sum m_\nu$.
Separately, for ease of comparison of our results with previous studies, in Appendix~\ref{Appendix:C} we consider cosmologies where $N_\mathrm{eff}$ is another free parameter of the fit, in both the extended $\Lambda$CDM model and in the self-interacting neutrino cosmology; however, we note that the self-interacting neutrino model considered here does not require a departure of $N_\mathrm{eff}$ from its Standard-Model value.

In our sampling runs, we assume broad flat priors on all cosmological parameters listed above, a log-flat prior on $G_\mathrm{eff}$, and a Gaussian prior on $\tau_\mathrm{reio} = 0.065 \pm 0.015$ as a stand-in for low-$\ell$ EE data\footnote{We test the difference between adding the full low-$\ell$ EE likelihood to our analysis and using the tau prior as a proxy in Appendix~\ref{Appendix:D}, and find no appreciable difference.} \citep{Kreisch_2022}.
We limit the upper bound of our prior on $\mathrm{log}_{10}(G_\mathrm{eff} \ \mathrm{MeV}^2)$ to $-0.5$, as the equations of motion become too stiff for \texttt{CLASS} to evolve at $\mathrm{log}_{10}(G_\mathrm{eff} \ \mathrm{MeV}^2) > -0.5$ \citep{Oldengott_2017}. 
We do not expect this choice to impact our results, as our chains show good convergence to best-fit values well within our chosen prior range. 
We extend our analyses to smaller neutrino self-couplings, sampling down to $\mathrm{log}_{10}(G_\mathrm{eff} \ \mathrm{MeV}^2) = -7$. This is essential because the LSS data probe scales down to $k \sim 1 \ h / \mathrm{Mpc}$, which are sensitive to interaction strengths as low as $\mathrm{log}_{10}(G_\mathrm{eff} \ \mathrm{MeV}^2) \sim -6$. 

In our \texttt{MultiNest} runs, we set the number of live points to 1000, the target sampling efficiency to 0.8, and the accuracy threshold for the log Bayesian evidence to 20\%, ensuring that our analysis is optimized for parameter estimation. We repeat a subset of our runs for 2000 live points to ensure that we do not miss any high-probability modes in the posterior, and we find no difference. We also check that in the cases where the posterior is not multi-modal, MCMC returns consistent results with nested sampling.

\begin{figure*}[t!]
\centering
\includegraphics[scale=0.9]{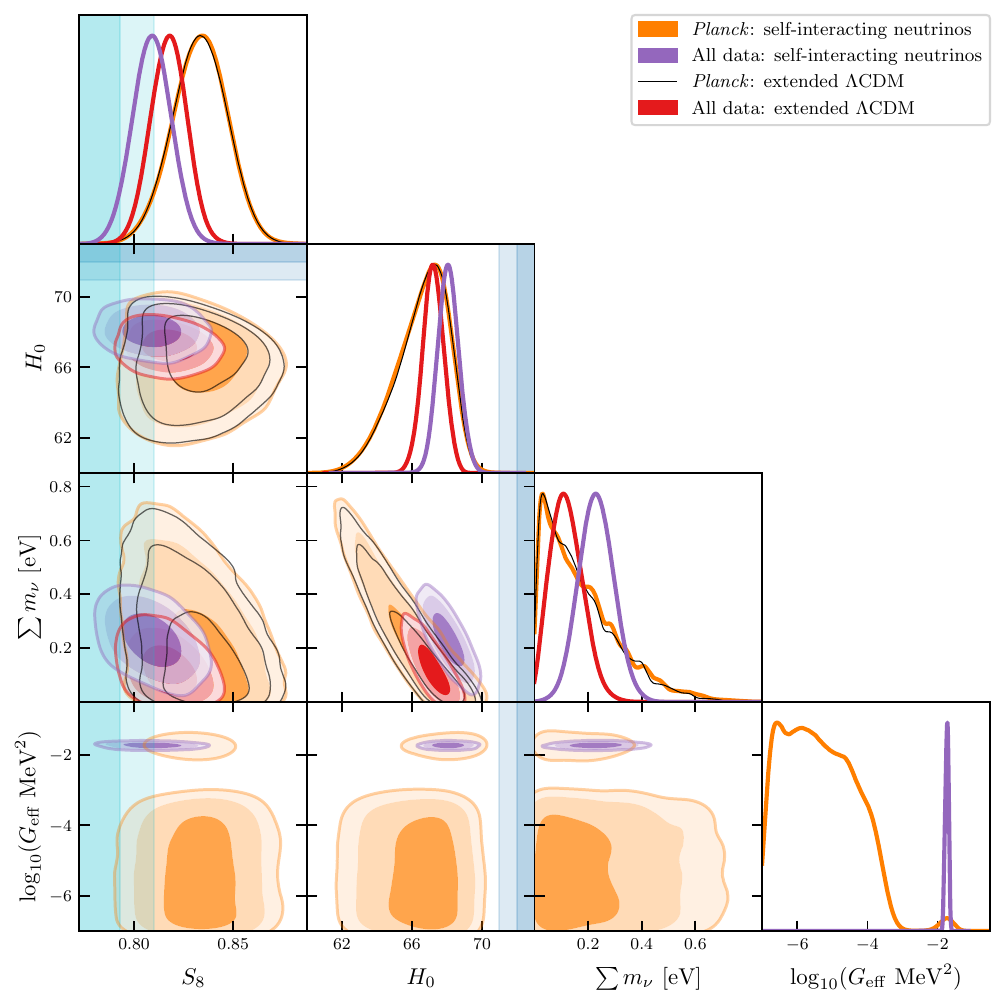}
\caption{68\%, 95\%, and 99\% confidence-level countours for a 2D marginalized posterior distribution of a subset of cosmological parameters for two models: the self-interacting neutrino model (referred to as $G_\mathrm{eff}$+$\sum m_\nu$ in the text) and the extended standard cosmology ($\Lambda$CDM+$\sum m_\nu$). We show the results of a \textit{Planck}-only analysis and of a combined analysis of \textit{Planck}, BOSS, Lyman-$\alpha$, and DES data (labeled as ``all data''), for both models; the former recovers the results of previous work, which the latter shows the key novel result of the present work. The blue and teal shaded bands show the $\mathrm{SH_0ES}$ measurement of $H_0$ and the DES prior on $S_8$, respectively. We note that the interacting-neutrino model mildly reduces both the Hubble tension and the $S_8$ tension between different data sets.
\label{fig:H0 S8 mnu}}
\end{figure*}

\section{\label{sec:results}Results\protect}
\begin{table*}[t!]
\centering
\caption{$\Delta\chi_{\mathrm{min}}^2$ for our baseline self-interacting neutrino model $G_\mathrm{eff}$+$\sum m_\nu$, compared to $\Lambda$CDM and an interaction-free extension of $\Lambda$CDM in which $\sum m_\nu$ is a free parameter ($\Lambda$CDM+$\sum m_\nu$). Rows present individual contributions from different subsets of data to the full analysis of \textit{Planck} + BOSS + Lyman-$\alpha$ + DES.} \label{tab:chi_SIv}
\begin{tabular}{|c|c|c|c|c|c|}
\hline
Data set & $\Delta\chi^2$ wrt $\Lambda$CDM+$\sum m_\nu$ & $\Delta\chi^2$ wrt $\Lambda$CDM \\
\hline 
\hline
\textit{Planck} low--$\ell$ TT & $-0.01$ & $+0.09$ \\ [0.5ex] 
\textit{Planck} 
 high--$\ell$ & $-0.90$ & $-1.52$ \\ [0.5ex] 
\textit{Planck} 
 lensing & $-0.08$ & $-0.18$ \\ [0.5ex] 
BOSS & $+0.38$ & $-1.53$ \\ [0.5ex] 
Lyman--$\alpha$ & $-24.91$ & $-26.02$ \\ [0.5ex] 
DES & $-2.78$ & $-1.03$ \\ [0.5ex] 
$\tau$ prior & $-0.14$ & $+0.18$ \\ [0.5ex]
\hline
Total & $-28.44$ & $-30.01$ \\
[0.5ex]
 \hline
\end{tabular}
\end{table*}
\raggedbottom
In Fig.~\ref{fig:H0 S8 mnu}, we show the 2D posterior probability distributions for key cosmological parameters, obtained from two different data sets: \textit{Planck} only (orange and black) and a combination of all data described in Sec.~\ref{sec:data} (\textit{Planck} + BOSS + Lyman-$\alpha$ + DES). We analyze each data set under two different cosmological models: the self-interacting neutrino cosmology ($G_\mathrm{eff}$+$\sum m_\nu$; orange and red), and the extended standard cosmology ($\Lambda$CDM+$\sum m_\nu$; black and gray).
All the results are derived assuming $N_\mathrm{eff}=3.046$ fixed. 

In the \textit{Planck}--only analysis of $G_\mathrm{eff}$+$\sum m_\nu$, we find that the posterior probability distribution is bimodal, consistent with the results from previous literature \cite{Kreisch_2022}. While the weak neutrino self-coupling (with a negligible delay in neutrino free-streaming) is statistically preferred in this case, a strong coupling is also consistent with the \textit{Planck} data. However, when the LSS likelihoods are included, the posterior mode consistent with $\Lambda$CDM is disfavored, and only the strongly-coupled mode of the posterior persists; see the bottom right panel of Fig.~\ref{fig:H0 S8 mnu}. The peak of the 1D marginalized posterior is at $\mathrm{log}(G_\mathrm{eff} \ \mathrm{MeV}^2)=-1.73_{-0.1}^{+0.09}$, corresponding to a delay of free-streaming until $z\sim 8300$, where the uncertainty captures the 95\% credible interval. In other words, the inclusion of the LSS data leads to a drastic increase in the significance of strong neutrino self-coupling, as compared to the \textit{Planck}--only analysis. 

In order to understand whether this preference is driven by the effects of the neutrino mass, or by the effects of neutrino self-interactions on matter clustering, we compare the quality of the fit for the interacting neutrino model to the quality of the fit for the standard $\Lambda$CDM model, as well as an extended $\Lambda$CDM+$\sum m_\nu$ model (where the the sum of neutrino masses is an additional free parameter). We find that the addition of neutrino self-interactions, rather than the sum of the neutrino masses, drives the improvement in the fit. Concretely, comparing $\chi^2$ of the self-interacting model to that of $\Lambda$CDM and $\Lambda$CDM+$\sum m_\nu$, we find a difference of $\Delta\chi_{\mathrm{}}^2 = -30.01$ and $-28.44$, respectively. This translates to a 5.3$\sigma$ preference for strong neutrino self-coupling, as compared to the standard cosmology. 

Table~\ref{tab:chi_SIv} displays the breakdown of $\Delta\chi^2$ contributions from each data set. 
The Lyman-$\alpha$ forest data dominate the preference toward the neutrino self-interacting model over $\Lambda$CDM, with DES adding a sub-dominant contribution; the rest of the data sets are largely agnostic in this sense. 
This preference can be understood in the context of modifications to the linear matter power spectrum $P(k)$, caused by neutrino self-scattering. 
As discussed in Sec.~\ref{sec:level1}, the interactions lead to a bump-like feature at $k\sim0.2 \ h/\mathrm{Mpc}$, and a subsequent suppression at smaller scales, altering both the amplitude and the slope of $P(k)$ in a scale-dependent manner; see Figs.~\ref{fig:power spectra} and \ref{fig:derivative}. Combined with the lower $A_\mathrm{s}$ and $n_\mathrm{s}$ values preferred in this model, the neutrino self-interaction produces a suppression at the
pivot scale $k_p \sim 1 \ h/\mathrm{Mpc}$, better fitting the Lyman-$\alpha$ measurements. At the same time, the scale-dependent suppression of power is also mildly favored by the DES prior on $S_8$, adding to the overall quality of the fit. 

As noted in \cite{Kreisch_2020, Kreisch_2022}, there are significant parameter degeneracies in the interacting-neutrino model; in many instances, they result in reconciliation of the best-fit parameter values inferred from the CMB, LSS, and other probes. For example, we confirm that the $G_\mathrm{eff}$+$\sum m_\nu$ model marginally eases the $H_0$ tension between the CMB + LSS and the supernova measurements from Supernova $H_0$ for the Equation of State ($\mathrm{SH_0ES}$) survey~\citep{Riess_2016,Riess:2021jrx}, as shown in Fig.~\ref{fig:H0 S8 mnu}. Specifically, the tension between $\mathrm{SH_0ES}$ and \textit{Planck} + BOSS + Lyman-$\alpha$ + DES is at 4.8$\sigma$ under the $\Lambda$CDM+$\sum m_\nu$ model, but it reduces to 4.1$\sigma$ under the self-interacting neutrino model. This indicates that a combined analysis of early and late time probes of the expansion may show an even stronger preference for the $G_\mathrm{eff}$+$\sum m_\nu$ model in comparison to $\Lambda$CDM+$\sum m_\nu$.
At the same time, we see a minor reduction in the $S_8$ tension between DES and \textit{Planck} + BOSS + Lyman-$\alpha$ + DES data; the tension is eased from 2.2$\sigma$ in $\Lambda$CDM+$\sum m_\nu$ to 1.7$\sigma$ when self-interactions are added. However, we note that we have included a prior on $S_8$ from DES in our analysis, pulling our results towards the DES value for $S_8$. A dedicated future analysis is thus needed to fully understand the impact of the neutrino interactions on the $S_8$ tension.

It is interesting to note that allowing for neutrino self-scattering also leads to a $3.5\sigma$ preference for a non-zero sum of neutrino masses, with the maximum of the marginalized 1D posterior at $\sum m_\nu=0.23\pm0.13$ eV at 95\% confidence, consistent with the bounds from neutrino oscillation experiments~\citep{Esteban:2018azc}, and in agreement with previous \textit{Planck} collaboration bounds \citep{Planck2018_VI}. 
This finding is also consistent with the $>2\sigma$ preference for non-vanishing neutrino mass found in previous \textit{Planck}--only analyses of self-interacting neutrinos in~\cite{Kreisch_2020}; the preference for non-vanishing neutrino mass arises from the need to suppress the amplitude of matter fluctuations at late times, as measured by CMB lensing in their analysis. It is thus unsurprising that, when analyzed with more LSS data sets which prefer a lower amplitude of matter fluctuations at late times, this model finds an even greater preference for non-vanishing neutrino mass. We also note a $\sim 2\sigma$ preference for a non-zero sum of neutrino masses in our all-data analysis even for the $\Lambda$CDM+$\sum m_\nu$ model ($\sum m_\nu=0.119^{+0.118}_{-0.111}$ eV at 95\% confidence).

We list the full set of constraints on all cosmological parameters for a \textit{Planck} + BOSS + Lyman-$\alpha$ + DES analysis of the $G_\mathrm{eff}$+$\sum m_\nu$ model in Appendix~\ref{Appendix:A}, and we show full posterior distributions for the $G_\mathrm{eff}$+$\sum m_\nu$ model in Appendix~\ref{Appendix:B}. In Appendix~\ref{Appendix:C}, we repeat our key analyses with the effective number of relativistic species $N_\mathrm{eff}$ as an additional free parameter in the fit. We note that the self-interacting neutrino model we consider in this work does not require deviation from the standard value of $N_\mathrm{eff}$ by default; however, for ease of comparison of our results with previous literature, and to explore parameter degeneracies with commonly-considered extensions of the standard cosmological model, we provide this analysis as well.

\section{\label{sec:discussion}Discussion and Conclusions\protect}

We considered a cosmological scenario in which flavor-universal neutrino self-interactions through a massive mediator particle, occurring in the early universe, delay the onset of neutrino free-streaming. We used the EFT of LSS to model matter distribution on mildly non-linear scales within the self-interacting neutrino cosmology for the first time. This enabled us to perform the first cosmological search for evidence of neutrino self-interactions using a combination of LSS data, including BOSS, eBOSS, and DES, along with the CMB measurements from \textit{Planck}. While previous CMB--only analyses found consistency with weak neutrino self-coupling, we find that the full data set strongly favors presence of neutrino self-scattering. In other words, we recover a single-mode posterior probability distribution for the neutrino self-coupling constant $G_\mathrm{eff}$, peaked at $\mathrm{log}(G_\mathrm{eff} \ \mathrm{MeV}^2)=-1.73_{-0.1}^{+0.09}$ at 95\% confidence. The LSS data contributes to a substantial $\sim 5\sigma$ preference for the strongly-coupled mode over $\Lambda$CDM, when all the data are analyzed in combination; this preference is driven by Lyman-$\alpha$ and, to a lesser extent, DES. The reported preference persists even when the sum of neutrino masses is allowed as a free parameter within the $\Lambda$CDM cosmology. Interestingly, we find that the combination of all data shows a preference for a non-vanishing sum of neutrino masses, at a level consistent with previous results from \textit{Planck}. Furthermore, the delay in neutrino free-streaming lowers the amplitude of matter fluctuations in the late universe, marginally easing the $S_8$ tension between LSS and CMB. Finally, neutrino self-scattering also decreases the size of the sound horizon, easing the tension between the early universe measurements of $H_0$ and the supernova data. Our key result is shown in Fig.~\ref{fig:H0 S8 mnu}.

The success of the interacting-neutrino model in fitting all the data simultaneously arises from subtle degeneracies in various standard cosmological parameter values, which all combine to give rise to the characteristic scale-dependence of the linear matter power spectrum, shown in Figs.~\ref{fig:power spectra} and \ref{fig:derivative}. However, the data that drive the preference for self-interactions are not sensitive to the detailed scale-dependence of $P(k)$. Namely, the Lyman-$\alpha$ forest traces the amplitude and the slope of $P(k)$ in a narrow range of scales around $k\sim 1 \ h/\mathrm{Mpc}$, while DES data is sensitive to the $P(k)$ amplitude around $k\sim 0.2 \ h/\mathrm{Mpc}$; finally, \textit{Planck} has an integrated sensitivity to a broad range of larger scales. Our results can thus be more broadly interpreted as an indication of a scale-dependent alteration in $P(k)$, which resembles the broad features produced by self-interacting neutrinos.

We further emphasize that the approach we adopted in this study was to take all data sets at face value when evaluating the viability of the neutrino self-interactions. However, it is possible that the likelihoods associated with the LSS or the CMB data 
are affected by unknown systematic effects, which may bias our results. We especially note that the reported preference for non-vanishing self-interaction is sensitive to the accuracy of the power spectrum reconstruction from Lyman-$\alpha$ measurements.
Indeed, tensions between 
the BOSS/eBOSS Lyman-$\alpha$ data and the \textit{Planck} $\Lambda$CDM model were previously reported~\citep{Palanque-Delabrouille:2015pga,Chabanier:2018rga,Hooper:2022byl}, but were not explored with a full combination of LSS and CMB data. 
Further insight may come 
from alternative approaches 
to the modeling of the Lyman-$\alpha$ forest 
power spectrum~\citep{Ivanov:2023yla},
as well as from its interpretation within new-physics models~\citep{Hooper:2022byl,rogers20245}.

The results presented in this study call for further investigation in several additional directions. First, self-scattering neutrinos suppress structure on a large range of scales, potentially affecting other observables beyond those considered in this study. For example, the Milky Way satellite galaxy census currently limits the suppression of power to $\sim 30\%$ compared to $\Lambda$CDM \citep{Nadler_2019,Nadler_2021}, at $k \sim 30 \ h/\mathrm{Mpc}$. Forthcoming galaxy surveys with Vera C. Rubin Observatory \citep{VeraRubin2018,Nadler_2019,Nadler_2021} and other facilities \citep{DESI:2016fyo,SPHEREx:2014bgr} will tighten this uncertainty further~\citep{Chudaykin:2019ock,Sailer:2021yzm,Nadler_2019, drlicawagner2019}, putting pressure on all beyond-CDM models that alter the matter power spectrum on small scales. The decrement of power appearing in the self-interacting neutrino model may also lead to signals at the level of precision projected for the Simons Observatory measurements \citep{SimonsObservatory:2018koc}. A holistic consideration of the growth of structure in massive interacting-neutrino cosmologies and its implications for upcoming observations is also warranted \citep{Nguyen_2023}. In addition, neutrino self-interactions require shifts in various cosmological parameters, including the sum of neutrino masses, in order to retain a good fit to the data. This implies that high-precision cosmological searches for $\sum m_\nu$ will provide critical information about neutrino interactions as well.

On the theory side, the preference for a delayed onset of neutrino free-streaming calls for a consideration of different types of particle interactions that could cause this delay. Different models for the interactions are constrained by experiments and astrophysical observations~\citep{PhysRevD.36.2895, Manohar:1987ec, Dicus:1988jh, Davoudiasl_2005, Sher_2011, Fayet_2006, PhysRevD.42.293, Blennow_2008, Galais_2012, Kachelriess_2000, Farzan_2003, Zhou_2011, Jeong_2018, chang2022powerful, Ahlgren_2013, Huang_2018, Venzor_2021, Ng_2014, Ioka_2014, cherry2016shortbaseline, Bilenky:1994ma, BARDIN1970121, bilenky1999secret, Brdar_2020, Lyu_2021, Lessa_2007, Bakhti_2017, Arcadi_2019, Blinov_2019, Damiano_1, Damiano_2, Damiano_3}, and a flavor-universal coupling where all three standard neutrinos interact and experience a delay in free-streaming may not be consistent with laboratory bounds, under this specific interaction model. On the other hand, laboratory experiments are already probing neutrino interaction physics at relevant levels \citep{Lyu_2021,Brinckmann_2021,Blinov_2019} and a flavor-specific neutrino self-interaction may be consistent with laboratory data and with cosmological observations. A dedicated analysis combining laboratory results with cosmological searches for neutrino self-scattering is also timely. 

\section*{Acknowledgements}
We gratefully acknowledge the support from the \textit{explore} Astrophysics fund for visiting scholars at USC. RA and VG acknowledge the support from NASA through the Astrophysics Theory Program, Award Number 21-ATP21-0135. 
VG acknowledges the support from the National Science Foundation (NSF) CAREER Grant No. PHY-2239205.
Exploration of the parameter space in this work was done with extensive use of \texttt{BSAVI}\footnote{\url{https://github.com/wen-jams/bsavi/tree/main}}.

\clearpage
\appendix
\onecolumngrid

\section{Full cosmological parameter constraints}
\label{Appendix:A}

In Table~\ref{tab:baseline constraints}, we show the full set of cosmological parameter constraints for a \textit{Planck} + BOSS + Lyman-$\alpha$ + DES analysis of the self-interacting neutrino model $G_\mathrm{eff}$+$\sum m_\nu$. In Table~\ref{tab:baseline constraints CDM}, we show the full set of cosmological parameter constraints for a \textit{Planck} + BOSS + Lyman-$\alpha$ + DES analysis of the $\Lambda$CDM+$\sum m_\nu$ model. 

\begin{table*}[htb!]
\centering
\caption{Full parameter constraints for a \textit{Planck} + BOSS + Lyman-$\alpha$ + DES analysis of our baseline self-interacting neutrino model $G_\mathrm{eff}$+$\sum m_\nu$. Bounds for standard cosmological parameters are given in the top half of the table, and bounds on \textit{Planck} and EFT bias parameters are given in the bottom half. The maximum of the full posterior is labeled ``Best-fit'', and the maxima of the marginalized posteriors are labeled ``Marginalized max''.
The superscripts (1), (2), (3), (4) 
of 
the galaxy bias parameters
$b_1,b_2,b_{\mathcal{G}_2}$
signify respectively the NGC $z=0.61$, SGC $z=0.61$,
NGC $z=0.38$, SGC $z=0.38$ BOSS DR12 data chunks. 
} 
\label{tab:baseline constraints}
\begin{tabular}{|l|c|c|c|c|}
 \hline
Parameter & Best-fit & Marginalized max $\pm \ \sigma$ & 95\% lower & 95\% upper \\ \hline
$100~\omega{}_\mathrm{b }$ &$2.272$ & $2.271\pm0.013$ & $2.245$ & $2.297$ \\
$\omega{}_\mathrm{c }$ &$0.1192$ & $0.1191\pm0.0009$ & $0.1174$ & $0.1209$ \\
$100~\theta{}_\mathrm{s }$ &$1.0467$ & $1.0466_{-0.00037}^{+0.00036}$ & $1.0458$ & $1.0473$ \\
$\mathrm{ln}(10^{10}A_\mathrm{s })$ &$3.018$ & $3.02\pm0.022$ & $2.976$ & $3.065$ \\
$n_\mathrm{s }$ &$0.9447$ & $0.9448\pm0.0033$ & $0.9382$ & $0.9513$ \\
$\tau{}_\mathrm{reio }$ &$0.0743$ & $0.07414_{-0.0118}^{+0.0119}$ & $0.05029$ & $0.09824$ \\
$\mathrm{log}_{10}(G_\mathrm{eff} \ \mathrm{MeV}^2)$ & $-1.722$ & $-1.731_{-0.045}^{+0.055}$ & $-1.827$ & $-1.645$ \\
$\sum m_\nu \ [\mathrm{eV}]$ & $0.211$ & $0.23_{-0.066}^{+0.065}$ & $0.104$ & $0.359$ \\
$z_\mathrm{reio }$ &$9.555$ & $9.518_{-1.066}^{+1.079}$ & $7.248$ & $11.61$ \\
$\Omega{}_{\Lambda }$ &$0.6905$ & $0.6884_{-0.0072}^{-0.0073}$ & $0.6741$ & $0.7027$ \\
$Y_\mathrm{He}$ &$0.248$ & $0.248\pm{5.5e-05}$ & $0.2479$ & $0.2481$ \\
$H_0$ &$68.255$ & $68.072_{-0.594}^{+0.588}$ & $66.925$ & $69.273$ \\ 
$10^{+9}A_\mathrm{s }$ &$2.044$ & $2.049_{-0.045}^{+0.046}$ & $1.961$ & $2.144$ \\
$\sigma_8$ &$0.7973$ & $0.794_{-0.0125}^{+0.0127}$ & $0.7702$ & $0.8187$ \\
$S_8$ &$0.81$ & $0.809\pm0.01$ & $0.79$ & $0.828$ \\
\hline
$A_\mathrm{planck}$ &$0.99913$ & $1.00038_{-0.00252}^{+0.00249}$ & $0.99541$ & $1.00533$ \\
$b^{(1)}_{1 }$ &$2.032$ & $2.041\pm0.048$ & $1.945$ & $2.137$ \\
$b^{(1)}_{2 }$ &$-0.5424$ &
$-0.4119_{-0.5919}^{+0.5326}$ & $-1.4523$ & $0.7166$ \\
$b^{(1)}_{{\mathcal{G}_2} }$ &$-0.4111$ & $-0.307_{-0.2819}^{+0.2804}$ & $-0.8695$ & $0.248$ \\
$b^{(2)}_{1 }$ &$2.224$ & $2.202_{-0.058}^{+0.057}$ & $2.087$ & $2.315$ \\
$b^{(2)}_{2 }$ &$-0.5069$ & $-0.3663_{-0.6239}^{+0.6126}$ & $-1.5416$ & $0.9082$ \\
$b^{(2)}_{{\mathcal{G}_2} }$ &$-0.0416$ & $-0.1446_{-0.3219}^{+0.3236}$ & $-0.7735$ & $0.5084$ \\
$b^{(3)}_{1 }$ &$1.968$ & $1.957_{-0.046}^{+0.047}$ & $1.866$ & $2.047$ \\
$b^{(3)}_{2 }$ &$-0.2203$ & $-0.1266_{-0.4915}^{+0.4879}$ & $-1.0708$ & $0.8535$ \\
$b^{(3)}_{{\mathcal{G}_2} }$ & $-0.3043$ & $-0.3383_{-0.2727}^{+0.2787}$ & $-0.8567$ & $0.1933$ \\
$b^{(4)}_{1 }$ &$1.981$ & $1.993\pm0.057$ & $1.879$ & $2.104$ \\
$b^{(4)}_{2 }$ &$0.0936$ & $-0.3868_{-0.5396}^{+0.5374}$ & $-1.4514$ & $0.7244$ \\
$b^{(4)}_{{\mathcal{G}_2} }$ &$-0.2312$ & $-0.366_{-0.3163}^{+0.3107}$ & $-0.9774$ & $0.2646$ \\
\hline
\end{tabular}
\end{table*}
\raggedbottom

\begin{table*}[htb!]
\centering
\caption{Full parameter constraints for a \textit{Planck} + BOSS + Lyman-$\alpha$ + DES analysis of the $\Lambda$CDM+$\sum m_\nu$ model. Bounds for standard cosmological parameters are given in the top half of the table, and bounds on \textit{Planck} and EFT bias parameters are given in the bottom half. The maximum of the full posterior is labeled ``Best-fit'', and the maxima of the marginalized posteriors are labeled ``Marginalized max''.
The superscripts (1), (2), (3), (4) 
of 
the galaxy bias parameters
$b_1,b_2,b_{\mathcal{G}_2}$
signify respectively the NGC $z=0.61$, SGC $z=0.61$,
NGC $z=0.38$, SGC $z=0.38$ BOSS DR12 data chunks. 
} 
\label{tab:baseline constraints CDM}
\begin{tabular}{|l|c|c|c|c|}
 \hline
Parameter & Best-fit & Marginalized max $\pm \ \sigma$ & 95\% lower & 95\% upper \\ \hline
$100~\omega{}_\mathrm{b }$ &$2.237$ & $2.244\pm0.013$ & $2.218$ & $2.269$ \\
$\omega{}_\mathrm{c }$ &$0.1197$ & $0.1192\pm0.0009$ & $0.1174$ & $0.1209$ \\
$100~\theta{}_\mathrm{s }$ &$1.0419$ & $1.0419\pm0.00029$ & $1.0413$ & $1.0425$ \\
$\mathrm{ln}(10^{10}A_\mathrm{s })$ &$3.05$ & $3.054_{-0.022}^{+0.023}$ & $3.01$ & $3.098$ \\
$n_\mathrm{s }$ &$0.9584$ & $0.9603_{-0.0032}^{+0.0033}$ & $0.9539$ & $0.9668$ \\
$\tau{}_\mathrm{reio }$ &$0.0557$ & $0.05969_{-0.0118}^{+0.0117}$ & $0.03662$ & $0.08363$ \\
$\sum m_\nu \ [\mathrm{eV}]$ & $0.147$ & $0.119_{-0.069}^{+0.053}$ & $0.008$ & $0.237$ \\
$z_\mathrm{reio }$ &$7.841$ & $8.174_{-1.162}^{+1.149}$ & $5.78$ & $10.421$ \\
$\Omega{}_{\Lambda }$ &$0.6771$ & $0.6835_{-0.0074}^{-0.0076}$ & $0.668$ & $0.698$ \\
$Y_\mathrm{He}$ &$0.2478$ & $0.2479_{-5.5e-05}^{+5.3e-05}$ & $0.2478$ & $0.248$ \\
$H_0$ &$66.714$ & $67.212_{-0.59}^{+0.607}$ & $66.006$ & $68.393$ \\ 
$10^{+9}A_\mathrm{s }$ &$2.111$ & $2.12\pm0.048$ & $2.028$ & $2.216$ \\
$\sigma_8$ &$0.7892$ & $0.796_{-0.011}^{+0.0126}$ & $0.7719$ & $0.8178$ \\
$S_8$ &$0.819$ & $0.8175\pm0.009$ & $0.799$ & $0.835$ \\
\hline
$A_\mathrm{planck}$ &$1.00098$ & $1.00002_{-0.00245}^{+0.0025}$ & $0.99517$ & $1.00493$ \\
$b^{(1)}_{1 }$ &$2.051$ & $2.036\pm0.048$ & $1.942$ & $2.131$ \\
$b^{(1)}_{2 }$ &$-0.4228$ &
$-0.5467_{-0.5566}^{+0.5506}$ & $-1.607$ & $0.5817$ \\
$b^{(1)}_{{\mathcal{G}_2} }$ &$-0.3297$ & $-0.42_{-0.2814}^{+0.2817}$ & $-0.9706$ & $0.133$ \\
$b^{(2)}_{1 }$ &$2.21$ & $2.179_{-0.058}^{+0.057}$ & $2.066$ & $2.293$ \\
$b^{(2)}_{2 }$ &$0.4694$ & $-0.4388_{-0.6845}^{+0.5879}$ & $-1.6424$ & $0.8124$ \\
$b^{(2)}_{{\mathcal{G}_2} }$ &$-0.1038$ & $-0.2336_{-0.3388}^{+0.3456}$ & $-0.9018$ & $0.4466$ \\
$b^{(3)}_{1 }$ &$1.97$ & $1.94\pm0.046$ & $1.851$ & $2.031$ \\
$b^{(3)}_{2 }$ &$0.0871$ & $-0.196_{-0.5232}^{+0.4773}$ & $-1.1501$ & $0.814$ \\
$b^{(3)}_{{\mathcal{G}_2} }$ & $-0.2893$ & $-0.4137_{-0.2831}^{+0.2838}$ & $-0.9674$ & $0.1417$ \\
$b^{(4)}_{1 }$ &$1.971$ & $1.979_{-0.056}^{+0.057}$ & $1.867$ & $2.089$ \\
$b^{(4)}_{2 }$ &$-0.4912$ & $-0.4373_{-0.5699}^{+0.559}$ & $-1.4906$ & $0.7024$ \\
$b^{(4)}_{{\mathcal{G}_2} }$ &$-0.5116$ & $-0.4285_{-0.3173}^{+0.3218}$ & $-1.0545$ & $0.2202$ \\
\hline
\end{tabular}
\end{table*}
\raggedbottom

\pagebreak

\clearpage

\section{The full posterior probability distribution}
\label{Appendix:B}

\begin{figure*}[h!]
\includegraphics[scale=0.66]{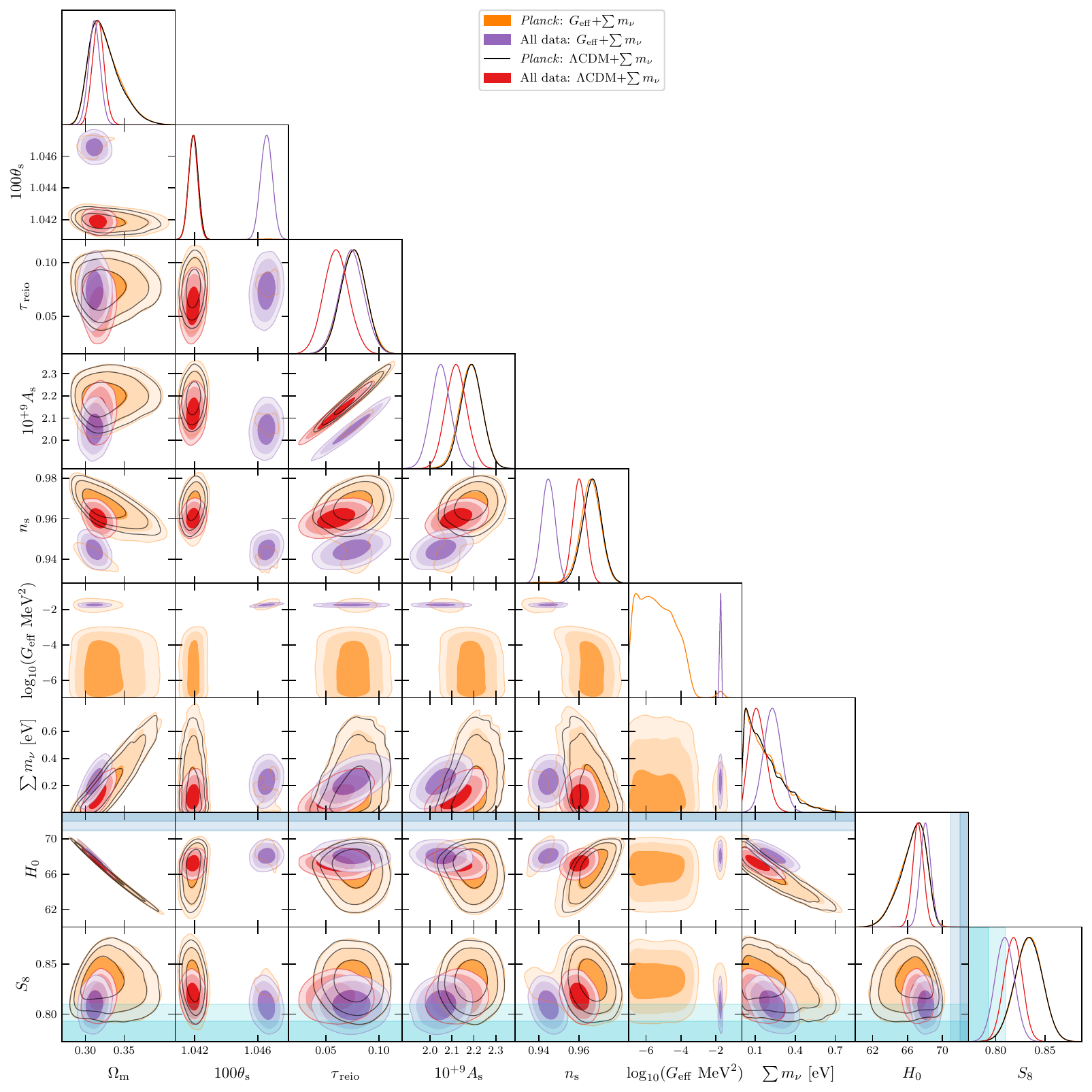}
\caption{68\%, 95\%, and 99\% confidence level marginalized posterior distributions of all cosmological parameters for $\Lambda$CDM+$\sum m_\nu$ (gray and black) and for the self-interacting neutrino model $G_\mathrm{eff}$+$\sum m_\nu$ (colored), from a combined analysis of \textit{Planck}, BOSS, Lyman-$\alpha$, and DES data (labeled as
``all data'')  and a \textit{Planck}-only analysis. The blue and teal shaded bands show the $\mathrm{SH_0ES}$ measurement of $H_0$ and the DES measurement of $S_8$, respectively.
\label{fig:posteriors}}
\end{figure*}

In Fig.~\ref{fig:posteriors}, we show the full marginalized posterior distributions for all relevant parameters in a \textit{Planck} + BOSS + Lyman-$\alpha$ + DES analysis of our baseline self-interacting neutrino model $G_\mathrm{eff}$+$\sum m_\nu$, with $N_\mathrm{eff}$ fixed at 3.046. 

\raggedbottom

\pagebreak

\clearpage

\section{Implications of additional relativistic species}
\label{Appendix:C}

\begin{figure*}[h!]
\centering
\includegraphics[scale=0.3]{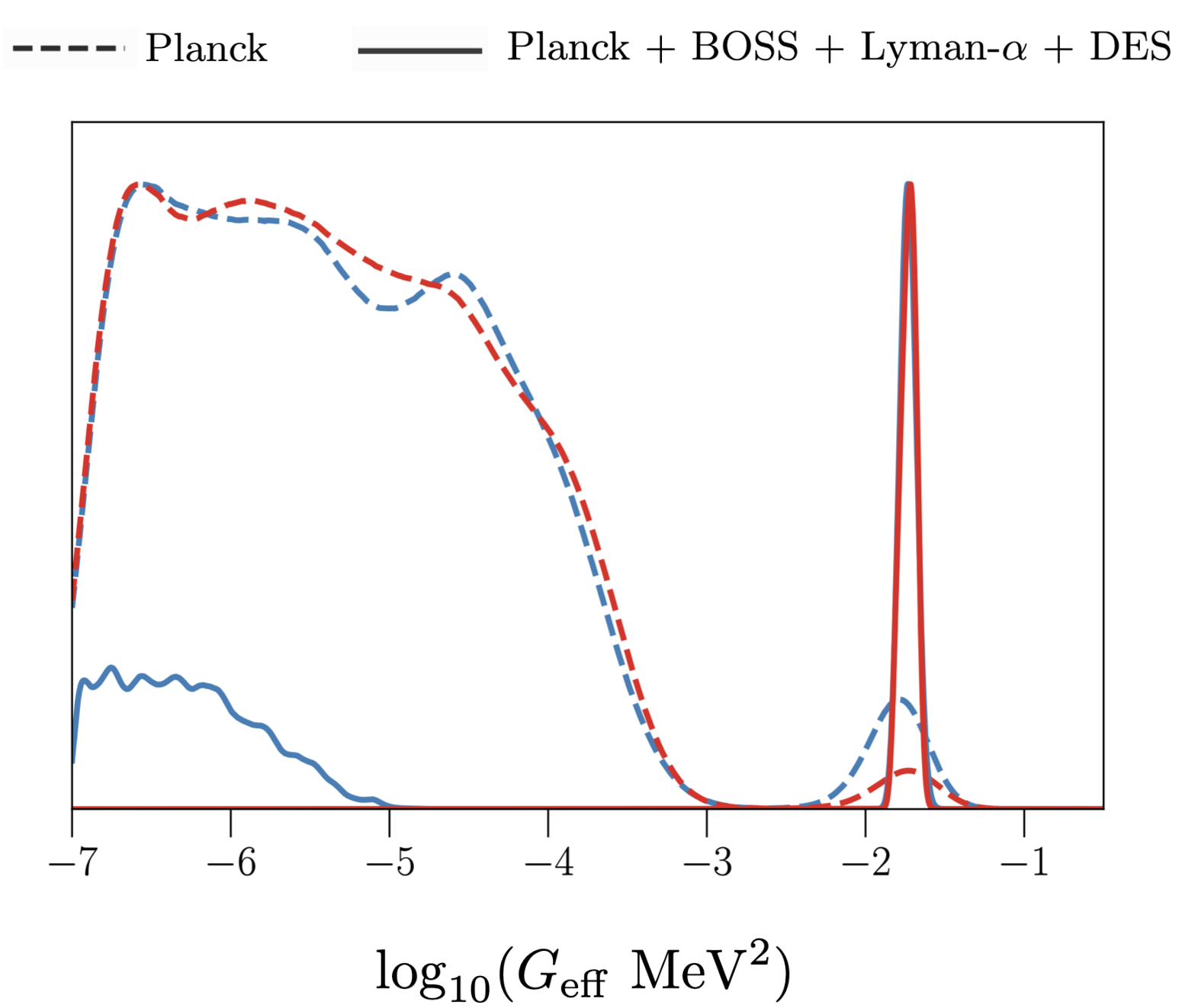}
\caption{1D marginalized posterior distribution for the neutrino self-interaction coupling parameter $G_\mathrm{eff}$. We show the posterior derived from a joint analysis of \textit{Planck}, BOSS, Lyman-$\alpha$, and DES data (solid) as well as the same posterior obtained from a \textit{Planck}-only analysis (dashed); the red corresponds to the analysis where $N_\mathrm{eff}=3.046$ is fixed at its standard model value, while the blue lines correspond to the case where it is a free parameter of the fit. In all four cases, the sum of the neutrino masses is a free parameter of the fit. We note that the addition of LSS data enhances the statistical significance of the strongly-interacting posterior mode in comparison to $\Lambda$CDM, regardless of whether or not $N_\mathrm{eff}$ is a free parameter. 
\label{fig:Geff with Neff}}
\end{figure*}

We now consider cosmologies that feature additional relativistic degrees of freedom. We emphasize that our main results focus on neutrino self-coupling as a method of reconciling the inconsistency between \textit{Planck} CMB data and LSS data. Our analysis in this section explores varying $N_\mathrm{eff}$ as an alternative way of reconciling this discrepancy. 
We label the corresponding extension of the standard cosmology as $\Lambda$CDM+$\sum m_\nu$+$N_\mathrm{eff}$ and the extension of the interacting-neutrino cosmology as $G_\mathrm{eff}$+$\sum m_\nu$+$N_\mathrm{eff}$.
In Fig.~\ref{fig:Geff with Neff}, we show the 1D posterior probability distribution for the self-coupling constant $G_\mathrm{eff}$, obtained from four different analyses: \textit{Planck}--only analyses of $G_\mathrm{eff}$+$\sum m_\nu$ and $G_\mathrm{eff}$+$\sum m_\nu$+$N_\mathrm{eff}$, and \textit{Planck} + BOSS + Lyman-$\alpha$ + DES analyses of $G_\mathrm{eff}$+$\sum m_\nu$ and $G_\mathrm{eff}$+$\sum m_\nu$+$N_\mathrm{eff}$. Regardless of the choice of additional free parameters, the inclusion of the LSS data leads to an increase in the significance of the strongly-interacting regime as compared to the \textit{Planck}--only analysis, but the relative significance is different depending on the chosen parameter space.

\begin{figure*}[h!]
\includegraphics[scale=0.56]{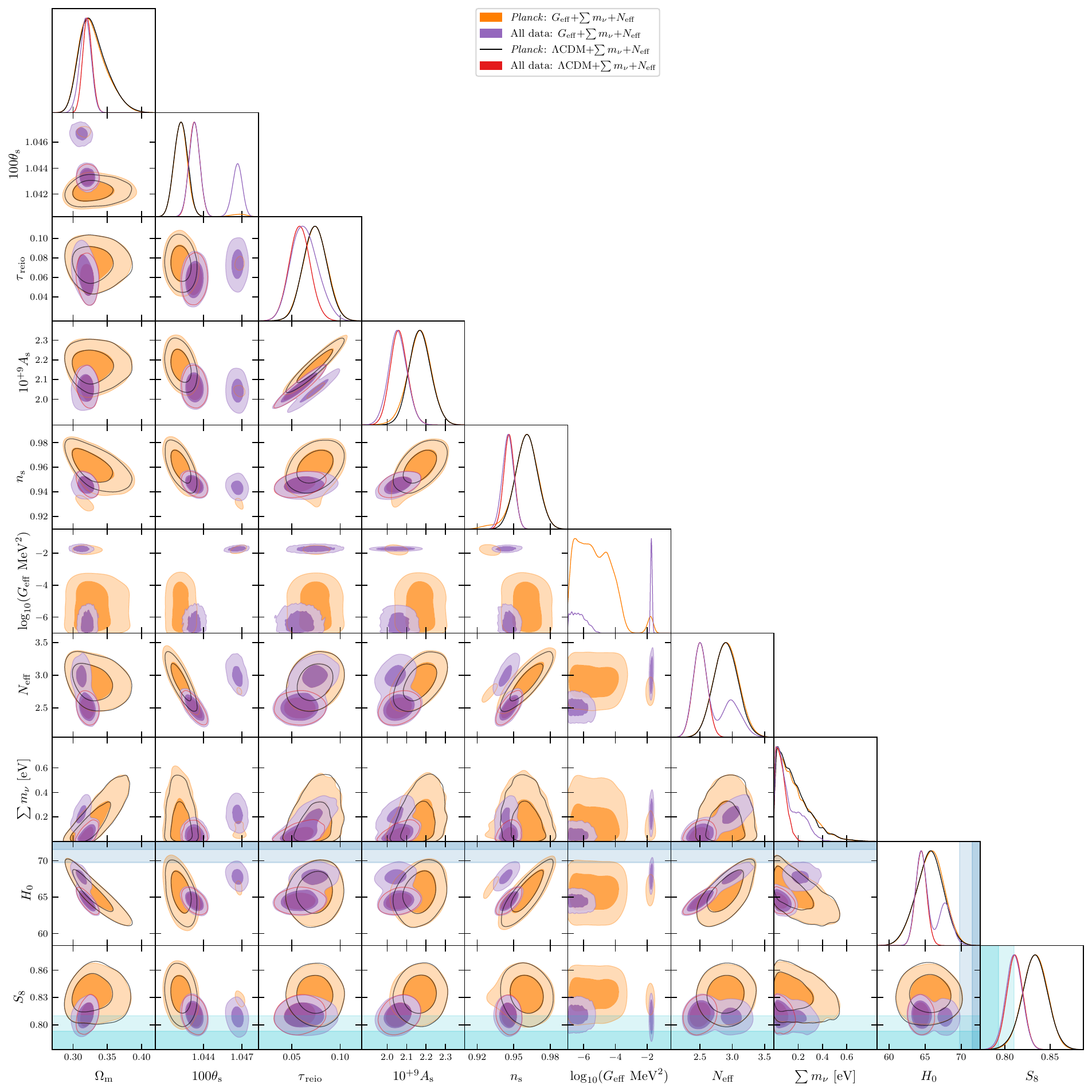}
\caption{68\% and 95\% confidence level marginalized posterior distributions of all cosmological parameters for $\Lambda$CDM+$\sum m_\nu$+$N_\mathrm{eff}$ (gray and black) and for the strongly-interacting mode of the $G_\mathrm{eff}$+$\sum m_\nu$+$N_\mathrm{eff}$ model (colored), from a combined analysis of \textit{Planck}, BOSS, Lyman-$\alpha$, and DES data (labeled as
``all data'')  and a \textit{Planck}-only analysis. The blue and teal shaded bands show the $\mathrm{SH_0ES}$ measurement of $H_0$ and the DES measurement of $S_8$, respectively.
\label{fig:freeing_Neff}}
\end{figure*}

As compared to the $\Lambda$CDM+$\sum m_\nu$+$N_\mathrm{eff}$ model, the $G_\mathrm{eff}$+$\sum m_\nu$+$N_\mathrm{eff}$ with $G_\mathrm{eff}$ consistent with the SI$\nu$ mode presents a mild improvement in fit, $\Delta\chi_{\mathrm{min}}^2 = -3.62$, corresponding to a 1.9$\sigma$ preference for strong neutrino self-interactions. In this scenario, however, the inferred mean value for $N_\mathrm{eff}$ under $\Lambda$CDM+$\sum m_\nu$+$N_\mathrm{eff}$ is $N_\mathrm{eff}=2.5\pm0.21$ at 95\% confidence, which is difficult to model in standard cosmology \citep{An_2022_Neff,dvorkin2022dark}. Table~\ref{tab:chi_free} displays the breakdown of $\Delta\chi^2$ contributions from each data set for the $G_\mathrm{eff}$+$\sum m_\nu$+$N_\mathrm{eff}$ model, as compared to $\Lambda$CDM+$\sum m_\nu$+$N_\mathrm{eff}$. We show full posterior distributions for a \textit{Planck} + BOSS + Lyman-$\alpha$ + DES analysis of the $G_\mathrm{eff}$+$\sum m_\nu$+$N_\mathrm{eff}$ model and the $\Lambda$CDM+$\sum m_\nu$+$N_\mathrm{eff}$ model in Fig.~\ref{fig:freeing_Neff}.

The low $N_\mathrm{eff}$ measured by $\Lambda$CDM+$\sum m_\nu$+$N_\mathrm{eff}$ can be understood in terms of the Lyman-$\alpha$ data's preference for a lower $A_\mathrm{s}$ and $n_\mathrm{s}$, which are degenerate with both $N_\mathrm{eff}$ and $G_\mathrm{eff}$. In order to fit both this data and the high-$\ell$ tail of the \textit{Planck} TT spectrum, the $\Lambda$CDM+$\sum m_\nu$+$N_\mathrm{eff}$ model significantly decreases $N_\mathrm{eff}$ from its standard value. This is contrasted with the $G_\mathrm{eff}$+$\sum m_\nu$+$N_\mathrm{eff}$ model's ability to toggle $G_\mathrm{eff}$ and thereby offset the effect of a lower $A_\mathrm{s}$ and $n_\mathrm{s}$ on the high-$\ell$ tail of the TT spectrum, without adjusting $N_\mathrm{eff}$. $N_\mathrm{eff}$ remains consistent with the three known neutrino species under the $G_\mathrm{eff}$+$\sum m_\nu$+$N_\mathrm{eff}$ model, with a mean value of $N_\mathrm{eff}=3^{+0.31}_{-0.27}$ at 95\% confidence; see Fig.~\ref{fig:freeing_Neff}. 

Since $N_\mathrm{eff}$ is highly degenerate with $H_0$, the $\Lambda$CDM+$\sum m_\nu$+$N_\mathrm{eff}$ model measures an $H_0$ value of $64.4 \pm 1.5$ at 95\% confidence, inflating the tension between $\mathrm{SH_0ES}$ and \textit{Planck} + BOSS + Lyman-$\alpha$ + DES for this model to 6.7$\sigma$. However, the SI$\nu$ mode of the $G_\mathrm{eff}$+$\sum m_\nu$+$N_\mathrm{eff}$ model retains an $H_0$ tension of 3.8$\sigma$ under this extended analysis. At the same time, we see no significant effect of the $G_\mathrm{eff}$+$\sum m_\nu$+$N_\mathrm{eff}$ model on the $S_8$ tension between DES and \textit{Planck} + BOSS + Lyman-$\alpha$ + DES data; the tension is only slightly reduced, from 1.83$\sigma$ in $\Lambda$CDM+$\sum m_\nu$+$N_\mathrm{eff}$ to 1.67$\sigma$ when self-interactions are added. This is indeed due to the $\Lambda$CDM+$\sum m_\nu$+$N_\mathrm{eff}$ model's lower measurement of $A_\mathrm{s}$ and $n_\mathrm{s}$, which decreases its measured value of $S_8$.

In Table~\ref{tab:Neff free constraints}, we show the full set of cosmological parameter constraints for a \textit{Planck} + BOSS + Lyman-$\alpha$ + DES analysis of the strongly-interacting mode in the $G_\mathrm{eff}$+$\sum m_\nu$+$N_\mathrm{eff}$ model. In Table~\ref{tab:Neff free constraints CDM}, we show the full set of cosmological parameter constraints for a \textit{Planck} + BOSS + Lyman-$\alpha$ + DES analysis of the $\Lambda$CDM+$\sum m_\nu$+$N_\mathrm{eff}$ model. Allowing for neutrino self-scattering with a varying $N_\mathrm{eff}$ again leads to a $>3\sigma$ preference for a non-zero sum of neutrino masses, with the mean value at $\sum m_\nu=0.22_{-0.14}^{+0.15}$ eV at 95\% confidence. We note no preference for a non-zero sum of neutrino masses in our all-data analysis of the $\Lambda$CDM+$\sum m_\nu$+$N_\mathrm{eff}$ model, as opposed to the $\sim 2 \sigma$ preference in the $\Lambda$CDM+$\sum m_\nu$ model discussed earlier.

\begin{table*}[b!]
\centering
\caption{$\Delta\chi_{\mathrm{min}}^2$ for the strongly-coupled mode (SI$\nu$) of the $G_\mathrm{eff}$+$\sum m_\nu$+$N_\mathrm{eff}$ model, compared to $\Lambda$CDM+$\sum m_\nu$+$N_\mathrm{eff}$. Rows present individual contributions from different subsets of data to the full analysis of \textit{Planck} + BOSS + Lyman-$\alpha$ + DES.} \label{tab:chi_free}
\begin{tabular}{|c|c|c|c|c|c|}
\hline
Data set & $\Delta\chi^2$ wrt $\Lambda$CDM+$\sum m_\nu$+$N_\mathrm{eff}$ 
\\
\hline 
\hline
\textit{Planck} low--$\ell$ TT & $-1.06$ \\ [0.5ex] 
\textit{Planck} high--$\ell$ & $-3.90$ \\ [0.5ex] 
\textit{Planck} lensing & $-0.08$ \\ [0.5ex] 
BOSS  & $-1.72$ \\ [0.5ex] 
Lyman--$\alpha$ & +1.23 \\ [0.5ex] 
DES & +2.31 \\ [0.5ex] 
$\tau$ prior & $-0.4$ \\ [0.5ex]
\hline
Total  & $-3.62$ \\
[0.5ex]
 \hline
\end{tabular}
\end{table*}
\raggedbottom

\begin{table*}[htb!]
\centering
\caption{Full parameter constraints for a \textit{Planck} + BOSS + Lyman-$\alpha$ + DES analysis of the strongly-interacting mode in the $G_\mathrm{eff}$+$\sum m_\nu$+$N_\mathrm{eff}$ model. Bounds for standard cosmological parameters are given in the top half of the table, and bounds on \textit{Planck} and EFT bias parameters are given in the bottom half. The maximum of the full posterior is labeled ``Best-fit'', and the maxima of the marginalized posteriors are labeled ``Marginalized max''.
The superscripts (1), (2), (3), (4) 
of 
the galaxy bias parameters
$b_1,b_2,b_{\mathcal{G}_2}$
signify respectively the NGC $z=0.61$, SGC $z=0.61$,
NGC $z=0.38$, SGC $z=0.38$ BOSS DR12 data chunks. 
} 
\label{tab:Neff free constraints}
\begin{tabular}{|l|c|c|c|c|}
 \hline
Parameter & Best-fit & Marginalized max $\pm \ \sigma$ & 95\% lower & 95\% upper \\ \hline
$100~\omega{}_\mathrm{b }$ &$2.254$ & $2.267_{-0.017}^{+0.018}$ & $2.233$ & $2.3$ \\
$\omega{}_\mathrm{c }$ &$0.1177$ & $0.1184_{-0.003}^{+0.002}$ & $0.1138$ & $0.1236$ \\
$100~\theta{}_\mathrm{s }$ &$1.0466$ & $1.0466_{-0.00036}^{+0.0004}$ & $1.0458$ & $1.0474$ \\
$\mathrm{ln}(10^{10}A_\mathrm{s })$ &$3.006$ & $3.017_{-0.025}^{+0.024}$ & $2.968$ & $3.065$ \\
$n_\mathrm{s }$ &$0.9387$ & $0.9438_{-0.0049}^{+0.0047}$ & $0.9345$ & $0.954$ \\
$\tau{}_\mathrm{reio }$ &$0.0663$ & $0.07345_{-0.0123}^{+0.0121}$ & $0.04913$ & $0.09671$ \\
$\mathrm{log}_{10}(G_\mathrm{eff} \ \mathrm{MeV}^2)$ & $-1.757$ & $-1.732\pm0.051$ & $-1.825$ & $-1.635$ \\
$N_\mathrm{eff}$ & $2.929$ & $3.003_{-0.159}^{+0.136}$ & $2.731$ & $3.312$ \\
$\sum m_\nu \ [\mathrm{eV}]$ & $0.162$ & $0.223_{-0.073}^{+0.071}$ & $0.085$ & $0.376$ \\
$z_\mathrm{reio }$ &$8.802$ & $9.442_{-1.053}^{+1.199}$ & $7.09$ & $11.543$ \\
$\Omega{}_{\Lambda }$ &$0.6893$ & $0.6881\pm0.0071$ & $0.6741$ & $0.7027$ \\
$Y_\mathrm{He}$ &$0.2463$ & $0.2474\pm0.002$ & $0.2435$ & $0.2515$ \\
$H_0$ &$67.607$ & $67.844_{-0.941}^{+0.864}$ & $66.11$ & $69.696$ \\ 
$10^{+9}A_\mathrm{s }$ &$2.02$ & $2.044\pm0.05$ & $1.946$ & $2.142$ \\
$\sigma_8$ &$0.8004$ & $0.7931_{-0.0118}^{0.0123}$ & $0.7693$ & $0.8166$ \\
$S_8$ &$0.815$ & $0.809\pm0.009$ & $0.79$ & $0.828$ \\
\hline
$A_\mathrm{planck}$ &$1.00176$ & $1.00044_{-0.00247}^{+0.00245}$ & $0.99538$ & $1.00544$ \\
$b^{(1)}_{1 }$ &$2.058$ & $2.04_{-0.05}^{+0.049}$ & $1.942$ & $2.136$ \\
$b^{(1)}_{2 }$ &$0.1562$ &
$-0.4059_{-0.5977}^{+0.5402}$ & $-1.4846$ & $0.7833$ \\
$b^{(1)}_{{\mathcal{G}_2} }$ &$-0.1448$ & $-0.3108_{-0.2765}^{+0.2808}$ & $-0.8518$ & $0.2444$ \\
$b^{(2)}_{1 }$ &$2.187$ & $2.204_{-0.058}^{+0.057}$ & $2.089$ & $2.323$ \\
$b^{(2)}_{2 }$ &$-0.5872$ & $-0.3692_{-0.6492}^{+0.5895}$ & $-1.5428$ & $0.8389$ \\
$b^{(2)}_{{\mathcal{G}_2} }$ &$-0.4588$ & $-0.1494_{-0.3422}^{+0.346}$ & $-0.8118$ & $0.5243$ \\
$b^{(3)}_{1 }$ &$1.938$ & $1.957_{-0.046}^{+0.047}$ & $1.865$ & $2.048$ \\
$b^{(3)}_{2 }$ &$0.0329$ & $-0.1265_{-0.4988}^{+0.4849}$ & $-1.0877$ & $0.8468$ \\
$b^{(3)}_{{\mathcal{G}_2} }$ & $-0.3404$ & $-0.3378_{-0.2785}^{+0.2714}$ & $-0.8894$ & $0.2206$ \\
$b^{(4)}_{1 }$ &$1.954$ & $1.995_{-0.058}^{+0.056}$ & $1.878$ & $2.11$ \\
$b^{(4)}_{2 }$ &$-1.009$ & $-0.3853_{-0.5537}^{+0.5546}$ & $-1.4376$ & $0.7778$ \\
$b^{(4)}_{{\mathcal{G}_2} }$ &$-0.5242$ & $-0.3641_{-0.3176}^{+0.3164}$ & $-0.979$ & $0.267$ \\
\hline
\end{tabular}
\end{table*}

\raggedbottom

\begin{table*}[htb!]
\centering
\caption{Full parameter constraints for a \textit{Planck} + BOSS + Lyman-$\alpha$ + DES analysis of the $\Lambda$CDM+$\sum m_\nu$+$N_\mathrm{eff}$ model. Bounds for standard cosmological parameters are given in the top half of the table, and bounds on \textit{Planck} and EFT bias parameters are given in the bottom half. The maximum of the full posterior is labeled ``Best-fit'', and the maxima of the marginalized posteriors are labeled ``Marginalized max''.
The superscripts (1), (2), (3), (4) 
of 
the galaxy bias parameters
$b_1,b_2,b_{\mathcal{G}_2}$
signify respectively the NGC $z=0.61$, SGC $z=0.61$,
NGC $z=0.38$, SGC $z=0.38$ BOSS DR12 data chunks. 
} 
\label{tab:Neff free constraints CDM}
\begin{tabular}{|l|c|c|c|c|}
 \hline
Parameter & Best-fit & Marginalized max $\pm \ \sigma$ & 95\% lower & 95\% upper \\ \hline
$100~\omega{}_\mathrm{b }$ &$2.2$ & $2.204\pm0.015$ & $2.174$ & $2.234$ \\
$\omega{}_\mathrm{c }$ &$0.1094$ & $0.1105\pm0.002$ & $0.1068$ & $0.1142$ \\
$100~\theta{}_\mathrm{s }$ &$1.0435$ & $1.0433\pm0.00041$ & $1.0425$ & $1.0441$ \\
$\mathrm{ln}(10^{10}A_\mathrm{s })$ &$3.018$ & $3.026_{-0.021}^{+0.022}$ & $2.985$ & $3.068$ \\
$n_\mathrm{s }$ &$0.9453$ & $0.9462\pm0.0043$ & $0.9376$ & $0.9547$ \\
$\tau{}_\mathrm{reio }$ &$0.05542$ & $0.0582_{-0.0108}^{+0.0107}$ & $0.03699$ & $0.08013$ \\
$N_\mathrm{eff}$ & $2.464$ & $2.502_{-0.107}^{+0.106}$ & $2.293$ & $2.711$ \\
$\sum m_\nu \ [\mathrm{eV}]$ & $0.042$ & $0.06_{-0.055}^{+0.021}$ & $>0$ & $0.141$ \\
$z_\mathrm{reio }$ &$7.636$ & $7.893_{-1.042}^{+1.048}$ & $5.726$ & $9.937$ \\
$\Omega{}_{\Lambda }$ &$0.6836$ & $0.6792\pm0.0067$ & $0.6656$ & $0.6923$ \\
$Y_\mathrm{He}$ &$0.2395$ & $0.2401\pm0.002$ & $0.2369$ & $0.2432$ \\
$H_0$ &$64.567$ & $64.443_{-0.754}^{+0.749}$ & $62.945$ & $65.951$ \\ 
$10^{+9}A_\mathrm{s }$ &$2.045$ & $2.061_{-0.05}^{+0.04}$ & $1.978$ & $2.151$ \\
$\sigma_8$ &$0.7835$ & $0.7844_{-0.0091}^{0.0104}$ & $0.7643$ & $0.8027$ \\
$S_8$ &$0.805$ & $0.811\pm0.009$ & $0.793$ & $0.828$ \\
\hline
$A_\mathrm{planck}$ &$0.99992$ & $0.99997_{-0.00246}^{+0.00251}$ & $0.99506$ & $1.00482$ \\
$b^{(1)}_{1 }$ &$2.039$ & $2.04_{-0.045}^{+0.046}$ & $1.948$ & $2.131$ \\
$b^{(1)}_{2 }$ &$-0.425$ &
$-0.5502_{-0.5985}^{+0.529}$ & $-1.6134$ & $0.623$ \\
$b^{(1)}_{{\mathcal{G}_2} }$ &$-0.3101$ & $-0.4401_{-0.2812}^{+0.2785}$ & $-1.0028$ & $0.1251$ \\
$b^{(2)}_{1 }$ &$2.179$ & $2.187_{-0.056}^{+0.057}$ & $2.077$ & $2.297$ \\
$b^{(2)}_{2 }$ &$-0.7274$ & $-0.4415_{-0.6756}^{+0.5996}$ & $-1.6304$ & $0.8637$ \\
$b^{(2)}_{{\mathcal{G}_2} }$ &$-0.5534$ & $-0.2565_{-0.3426}^{+0.3479}$ & $-0.9316$ & $0.4245$ \\
$b^{(3)}_{1 }$ &$1.929$ & $1.945\pm0.044$ & $1.858$ & $2.032$ \\
$b^{(3)}_{2 }$ &$-0.0705$ & $-0.1502_{-0.5272}^{+0.4776}$ & $-1.0947$ & $0.8723$ \\
$b^{(3)}_{{\mathcal{G}_2} }$ & $-0.2667$ & $-0.4169_{-0.2783}^{+0.278}$ & $-0.9594$ & $0.1389$ \\
$b^{(4)}_{1 }$ &$1.952$ & $1.989_{-0.054}^{+0.055}$ & $1.88$ & $2.094$ \\
$b^{(4)}_{2 }$ &$-0.8715$ & $-0.4155_{-0.5645}^{+0.5652}$ & $-1.4778$ & $0.7047$ \\
$b^{(4)}_{{\mathcal{G}_2} }$ &$-0.2636$ & $-0.4273_{-0.3171}^{+0.3217}$ & $-1.0412$ & $0.2062$ \\
\hline
\end{tabular}
\end{table*}

\raggedbottom

\raggedbottom

\pagebreak

\clearpage

\section{The effect of low-$\ell$ polarization data}
\label{Appendix:D}

\begin{figure*}[h!]
\includegraphics[scale=0.82]{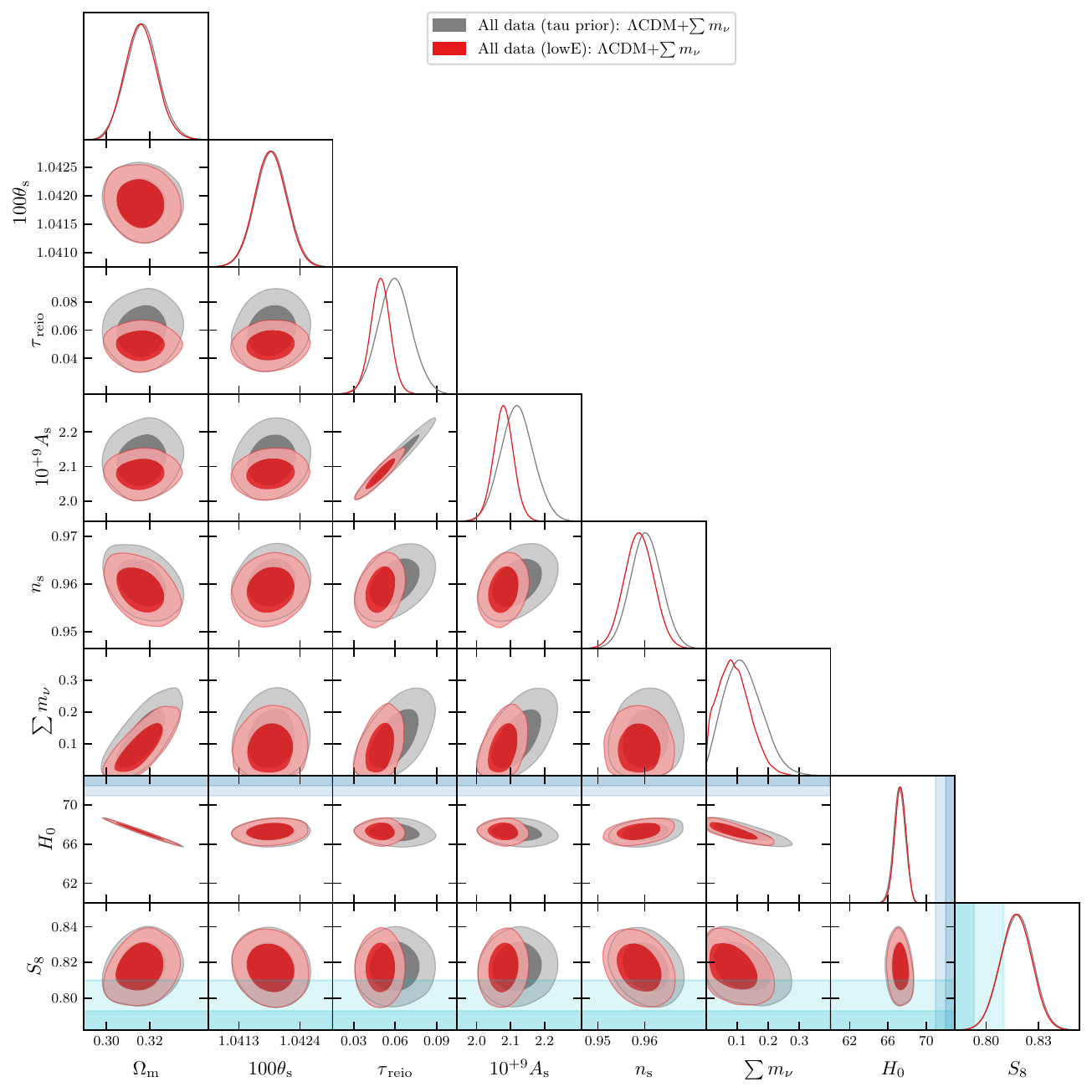}
\caption{68\% and 95\% confidence level marginalized posterior distributions of all cosmological parameters for $\Lambda$CDM+$\sum m_\nu$, from a combined analysis of \textit{Planck}, BOSS, Lyman-$\alpha$, and DES data with low-$\ell$ polarization data included (labeled as
``all data (lowE)'' in red) and a combined analysis of \textit{Planck}, BOSS, Lyman-$\alpha$, and DES data with a tau prior stand-in for low-$\ell$ polarization data (labeled as
``all data (tau prior)'' in gray). The blue and teal shaded bands show the $\mathrm{SH_0ES}$ measurement of $H_0$ and the DES measurement of $S_8$, respectively.
\label{fig:lowE_CDM}}
\end{figure*}

\begin{figure*}[h!]
\includegraphics[scale=0.67]{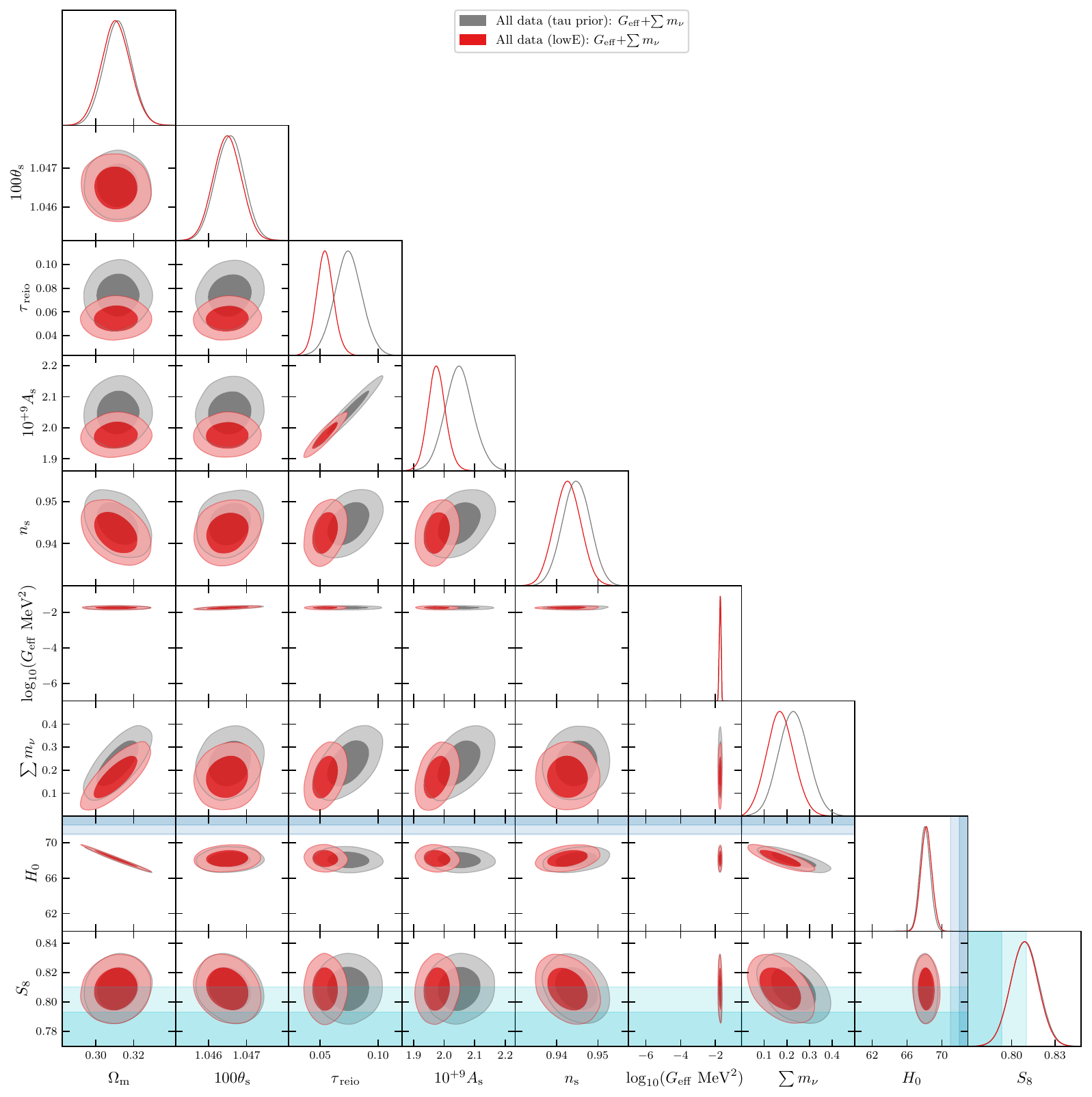}
\caption{68\% and 95\% confidence level marginalized posterior distributions of all cosmological parameters for $G_\mathrm{eff}$+$\sum m_\nu$, from a combined analysis of \textit{Planck}, BOSS, Lyman-$\alpha$, and DES data with low-$\ell$ polarization data included (labeled as
``all data (lowE)'' in red) and a combined analysis of \textit{Planck}, BOSS, Lyman-$\alpha$, and DES data with a tau prior stand-in for low-$\ell$ polarization data (labeled as
``all data (tau prior)'' in gray). The blue and teal shaded bands show the $\mathrm{SH_0ES}$ measurement of $H_0$ and the DES measurement of $S_8$, respectively.
\label{fig:lowE_SIv}}
\end{figure*}

We explore the impact of using the \textit{Planck} \texttt{SimAll} low-$\ell$ EE likelihood in our analysis, rather than the tau prior stand-in discussed in the main text. As reported in \cite{Natale_2020}, using a tau prior as a proxy for low-$\ell$ EE data leads to inflated values of $A_\mathrm{s}$, $n_\mathrm{s}$, and $\tau_\mathrm{reio}$ in a \textit{Planck} analysis of the $\Lambda$CDM model. We show posteriors for a \textit{Planck} + BOSS + Lyman-$\alpha$ + DES analysis of the $\Lambda$CDM+$\sum m_\nu$ model in Fig.~\ref{fig:lowE_CDM}, where the \textit{Planck} data includes all the aforementioned data as well as the low-$\ell$ EE likelihood. We also show posteriors for a \textit{Planck} + BOSS + Lyman-$\alpha$ + DES analysis of the $G_\mathrm{eff}$+$\sum m_\nu$ model in Fig.~\ref{fig:lowE_SIv}. Indeed, the inclusion of low-$\ell$ polarization data shifts the measured values of $A_\mathrm{s}$, $n_\mathrm{s}$, and $\tau_\mathrm{reio}$ to lower values, for both $\Lambda$CDM+$\sum m_\nu$ and $G_\mathrm{eff}$+$\sum m_\nu$. However, all of the other cosmological parameters remain the same as in the analysis where the tau prior is used as a stand-in for low-$\ell$ EE data, save for a minor shift in $\sum m_\nu$. In addition, the $G_\mathrm{eff}$+$\sum m_\nu$ model retains a 4.9$\sigma$ preference over the $\Lambda$CDM+$\sum m_\nu$ model with large-scale polarization data included, reiterating the $\sim 5 \sigma$ preference reported in the main text. We show in Table~\ref{tab:chi_lowE} the breakdown of $\Delta\chi^2$ contributions from each data set for the $G_\mathrm{eff}$+$\sum m_\nu$ model, as compared to $\Lambda$CDM+$\sum m_\nu$. We therefore conclude that the tau prior used in our analysis does not have any appreciable effect on our results.

\begin{table*}[b!]
\centering
\caption{$\Delta\chi_{\mathrm{min}}^2$ for the $G_\mathrm{eff}$+$\sum m_\nu$ model, compared to $\Lambda$CDM+$\sum m_\nu$. Rows present individual contributions from different subsets of data to the full analysis of \textit{Planck} + BOSS + Lyman-$\alpha$ + DES, which includes low-$\ell$ polarization data.} \label{tab:chi_lowE}
\begin{tabular}{|c|c|c|c|c|c|}
\hline
Data set & $\Delta\chi^2$ wrt $\Lambda$CDM+$\sum m_\nu$ 
\\
\hline 
\hline
\textit{Planck} low--$\ell$ TT & $-0.13$ \\ [0.5ex] 
\textit{Planck} low--$\ell$ EE & $+0.99$ \\ [0.5ex]
\textit{Planck} high--$\ell$ & $+0.15$ \\ [0.5ex] 
\textit{Planck} lensing & $-0.14$ \\ [0.5ex] 
BOSS  & $-1.12$ \\ [0.5ex] 
Lyman--$\alpha$ & $-22.18$ \\ [0.5ex] 
DES & $-1.87$ \\ [0.5ex] 
\hline
Total  & $-24.3$ \\
[0.5ex]
 \hline
\end{tabular}
\end{table*}
\raggedbottom

\raggedbottom

\raggedbottom

\pagebreak

\clearpage

\bibliography{apssamp}

\end{document}